\documentclass[acmsmall]{acmart}

\setcopyright{cc}
\setcctype{by}
\acmJournal{POMACS}
\acmYear{2026} \acmVolume{10}
\acmNumber{2} \acmArticle{39}
\acmMonth{6} \acmPrice{}
\acmDOI{10.1145/3805637}

\received{January 2026}
\received[revised]{March 2026}
\received[accepted]{April 2026}

\usepackage[camera]{dtrt}

\AtBeginDocument{%
  }
    
\usepackage{url}

\usepackage{breakurl}

\usepackage[nameinlink,noabbrev]{cleveref}
\usepackage{multirow}
\usepackage{makecell}
\usepackage{subcaption}
\usepackage{textcomp}
\usepackage{xcolor}
\usepackage{glossaries}
\usepackage{svg}
\usepackage{algorithm}
\usepackage{multirow}
\newacronym{pbs}{PBS}{Proposer-Builder Separation}
\newacronym{ofc}{OFC}{Order Flow Composition}
\newacronym{bft}{BFT}{Byzantine Fault Tolerant}
\newacronym{pos}{PoS}{Proof-of-Stake}
\newacronym{pow}{PoW}{Proof-of-Work}
\newacronym{mev}{MEV}{Maximal Extractable Value}
\newacronym[shortplural=DEXes]{dex}{DEX}{Decentralized Exchange}
\newacronym[shortplural=CEXes]{cex}{CEX}{Centralized Exchange} 
\newacronym{amm}{AMM}{Automated Market Maker}
\newacronym{defi}{DeFi}{Decentralized Finance}
\newacronym{tradfi}{TradFi}{Traditional Finance}
\newacronym{pga}{PGA}{Priority Gas Auction}
\newacronym{tob}{ToB}{Top-of-Block}
\newacronym{bob}{BoB}{Body-of-Block}
\newacronym{eob}{EoB}{End-of-Block}
\newacronym{ofa}{OFA}{Order Flow Auction}
\newacronym{0xce91228789b57deb45e66ca10ff648385fe7093b}{0xCe91228789B57DEb45e66Ca10Ff648385fE7093b}{MEV Blocker Rebates Safe}
\newacronym{rfq}{RFQ}{Request For Quote}
\newacronym{pm}{PM}{Profit Margin}
\newacronym{eoa}{EOA}{Externally Owned Account}
\newacronym{eof}{EOF}{Exclusive Order Flow}
\newacronym{msof}{MSOF}{Most Significant Order Flow}
\newacronym{jit}{JIT}{Just-In-Time}
\newacronym{lda}{LDA}{Linear Discriminant Analysis}
\newacronym{da}{DA}{Decoding Accuracy}
\newacronym{rpc}{RPC}{Remote Procedure Call}
\newacronym{brt}{BRT}{\texttt{beaverbuild}, \texttt{rysnc}, and \texttt{Titan}}
\newacronym{ep}{EP}{Exclusive Provider}
\newacronym{tee}{TEE}{Trusted Execution Environment}
\newacronym{il}{IL}{Inclusion List}
\newacronym{aps}{APS}{Attestor-Proposer Separation}
\newacronym{ea}{EA}{Execution Auction}
\newacronym{cr}{CR}{Censorship-Resistance}
\newacronym{hft}{HFT}{High-Frequency Trading}
\newacronym{l2}{L2}{Layer-2}
\newacronym{l1}{L1}{Layer-1}
\newacronym{bsc}{BSC}{Binance Smart Chain}
\newacronym{zk}{ZK}{Zero Knowledge}
\newacronym{mcp}{MCP}{Multiple Concurrent Proposers}
\newacronym{tvl}{TVL}{Total Value Locked}
\newacronym{sia}{SIA}{Sequence-Independent Arbitrage}
\newacronym{sda}{SDA}{Sequence-Dependent Arbitrage}
\newacronym{cdf}{CDF}{Cumulative Distribution Function}
\newacronym{ci}{CI}{Confidence Interval}
\newacronym{fp}{FP}{False Positive}
\newacronym{fn}{FN}{False Negative}
\newacronym{clob}{CLOB}{Central Limit Order Book}
\newacronym{p2p}{P2P}{Peer-to-Peer}
\newacronym{eip}{EIP}{Ethereum Improvement Proposal}
\newacronym{abm}{ABM}{Agent-Based Modeling}
\newacronym{fdv}{FDV}{Fully Diluted Valuation}
\newacronym{cpmm}{CPMM}{Constant Product Market Maker}
\newacronym{fcfs}{FCFS}{First-Come, First-Served}
\newacronym{epbs}{ePBS}{enshrined Proposer-Builder Separation}
\newacronym{se}{SE}{Simulation Experiment}
\newacronym{exp}{EXP}{Experiment}
\newacronym{hhi}{HHI}{Herfindahl–Hirschman Index}
\newcommand{\encircled}[2][0.9mm]{%
    \raisebox{.8pt}{%
        \textcircled{%
            \raisebox{0.35pt}{%
                \kern #1
                \scalebox{0.70}{#2}
            }%
        }%
    }%
}
\makeatletter
\newcommand*{\ensquared}[1]{\relax\ifmmode\mathpalette\@ensquared@math{#1}\else\@ensquared{#1}\fi}
\newcommand*{\@ensquared@math}[2]{\@ensquared{$\m@th#1#2$}}
\newcommand*{\@ensquared}[1]{%
\tikz[baseline,anchor=base]{\node[draw,outer sep=0pt,inner sep=0.6mm,minimum width=3.8mm] {#1};}} 
\makeatother

\usepackage{graphicx}
\usepackage{booktabs,arydshln}
\usepackage{siunitx}
\usepackage{subcaption}
\usepackage{ragged2e}
\usepackage{threeparttable}
\usepackage{siunitx}
\usepackage{tabularx,colortbl}
\usepackage{tikz}
\usepackage{enumitem}
\usepackage{adjustbox}
\usepackage{xspace}
\usepackage{longtable}
\usepackage{aliascnt}
\newtheorem{theorem}{Theorem}

\newaliascnt{proposition}{theorem}
\newtheorem{proposition}[proposition]{Proposition}
\aliascntresetthe{proposition}

\crefname{proposition}{proposition}{propositions}
\Crefname{proposition}{Proposition}{Propositions}
\makeatletter
\def\adl@drawiv#1#2#3{%
        \hskip.5\tabcolsep
        \xleaders#3{#2.5\@tempdimb #1{1}#2.5\@tempdimb}%
                #2\z@ plus1fil minus1fil\relax
        \hskip.5\tabcolsep}
\newcommand{\cdashlinelr}[1]{%
  \noalign{\vskip\aboverulesep
           \global\let\@dashdrawstore\adl@draw
           \global\let\adl@draw\adl@drawiv}
  \cdashline{#1}
  \noalign{\global\let\adl@draw\@dashdrawstore
           \vskip\belowrulesep}}
\makeatother
\glsdisablehyper

\makeatletter
\g@addto@macro\appendix{%
  \@ifundefined{chapterautorefname}{}{%
  }%
}
\makeatother
\AtBeginEnvironment{appendices}{%
}

\crefname{appendix}{Appendix}{Appendices}
\Crefname{appendix}{Appendix}{Appendices}
\pretocmd{\appendix}{%
  \crefalias{section}{appendix}%
  \crefalias{subsection}{appendix}%
  \crefalias{subsubsection}{appendix}%
}{}{}

\newtheorem*{remark}{Remark}
\newtheorem{lemma}{Lemma}
\crefname{lemma}{lemma}{lemmas}
\Crefname{lemma}{Lemma}{Lemmas}

\title{Geographical Centralization Resilience in Ethereum's Block-Building Paradigms}

\begin{document}

\author{Sen Yang}
\authornote{Work performed in part during an internship at Flashbots in 2025.}
\orcid{0000-0002-8866-2097}
\affiliation{%
  \institution{Yale University, IC3}
  \city{New Haven}
  \state{CT}
  \country{United States}}
\email{sen.yang@yale.edu}

\author{Burak Öz}
\orcid{0009-0003-7508-7112}
\affiliation{%
  \institution{Flashbots}
  \city{Munich}
  \state{Bayern}
  \country{Germany}}
\email{burak@flashbots.net}

\author{Fei Wu}
\authornote{Work performed in part during an internship at Flashbots in 2026.}
\orcid{0009-0004-5717-0219}
\affiliation{%
  \institution{King's College London}
  \city{London}
  \country{United Kingdom}}
\email{fei.wu@kcl.ac.uk}

\author{Fan Zhang}
\orcid{0000-0002-8525-4514}
\affiliation{%
  \institution{Yale University, IC3}
  \city{New Haven}
  \state{CT}
  \country{United States}}
\email{f.zhang@yale.edu}

\begin{abstract}
Decentralization has an important geographic dimension that conventional metrics, such as stake distribution, often overlook. Where validators operate affects resilience to regional shocks (e.g., outages, natural disasters, or government intervention) as well as fairness in reward access. Yet major blockchain protocols do not encode geographical location in their rules; instead, validator locations emerge from a combination of economic incentives, regulatory constraints, infrastructure availability, and validator deployment choices. When certain locations offer systematic advantages, validators may strategically co-locate to maximize expected rewards, as observed in Ethereum, where validators cluster along the Atlantic corridor, which exhibits favorable latency.

In this paper, we propose a formal model of validators' geographical positioning incentives under Ethereum's protocol design, capturing the interaction between its two block-building paradigms, local and external block building, and the geographical distribution of validators and information sources.
We analytically characterize the model under a mean-field approximation and complement this analysis with an agent-based simulation calibrated with real-world latency data to quantify how these incentives translate into geographical concentration under heterogeneous geographic and infrastructural conditions.

Our results show that Ethereum's block-building architecture is not geographically neutral. Both paradigms generate location-dependent payoffs and incentives to relocate closer to payoff-relevant parties in order to reduce propagation delays, although through different underlying mechanisms. Asymmetric access to information sources further amplifies geographical centralization. We also demonstrate that consensus parameters, such as attestation thresholds and slot times, modulate latency sensitivity and can amplify these effects, acting as protocol-level levers. Finally, we discuss the implications of our findings for protocol design and outline potential mitigation directions informed by our analysis.

\end{abstract}



\maketitle              %

\section{Introduction}
Decentralization is the core security assumption of permissionless blockchains, underpinning key properties such as integrity, safety, availability, and censorship resistance~\cite{nakamoto2008bitcoin,gencer2018decentralization,wahrstatter2024blockchain,yang2025decentralization}. These properties typically rely on the assumption that at least a fraction of participants behave honestly, or that conflicting incentives prevent collusion---that is, that control and behavior are not highly centralized or correlated. Consequently, assessing decentralization requires more than identifying who controls mining power or stake; it must also account for participants' geographical distribution.

\emph{Geographical decentralization} is particularly important for two reasons. First, security assumptions depend on participants not being overly correlated. If a large share of validators are co-located, they become jointly vulnerable to external shocks like natural disasters or power outages, as well as jurisdictional risks such as government intervention. A prominent example is the 2021 ban on Bitcoin mining in China, which triggered a sharp and sudden drop in global hash rate and forced miners to relocate at scale~\cite{thorn2021-china-bitcoin-ban}.
Second, geography affects fairness: validators farther from network hubs may face persistent latency disadvantages that reduce their ability to compete for rewards.

\begin{figure}[t]
  \centering
    \includegraphics[width=0.9\linewidth]{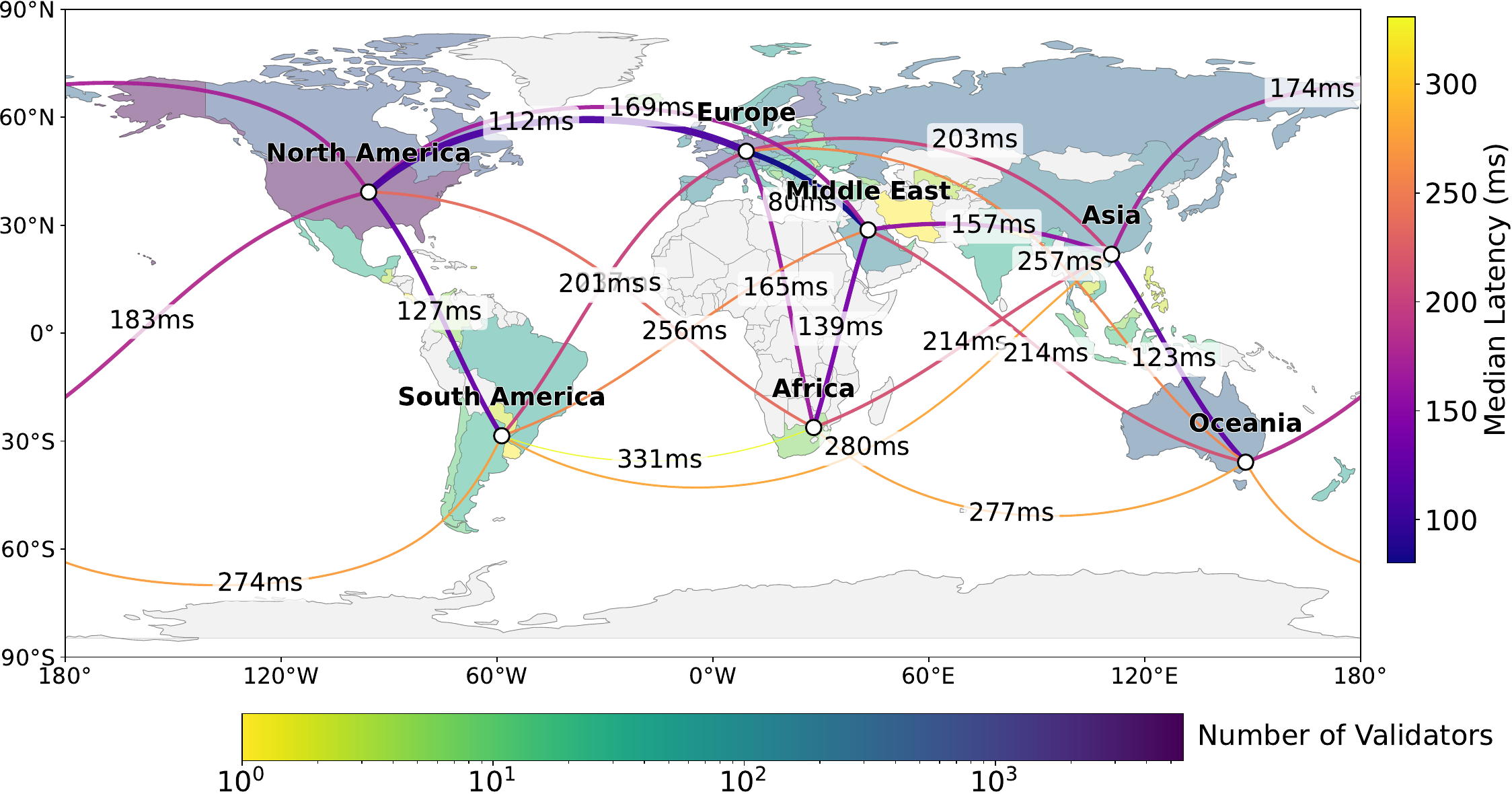}
  \caption{Validator distribution and inter-region Internet latencies illustrate the geographic concentration of Ethereum validators and the latency landscape across regions. Validator location data covers $\sim$\SI{16000}{} Ethereum validators, provided by Chainbound~\cite{chainbound_geovalidators}. Latency values are macro-regional medians of round-trip times between Google Cloud regions. Darker shading indicates a larger number of validators in a given country, showing that validators are primarily concentrated in the United States and Europe. Darker inter-regional links indicate lower average round-trip latency, highlighting the particularly low latency between North America and Europe relative to other regions.}
  \Description{}
  \label{fig:centralization101}
\end{figure}

Achieving geographical decentralization, however, is \emph{non-trivial}.
In major blockchains, participants' geographical locations are not encoded in the protocol rules; instead, they emerge from a combination of economic incentives, regulatory constraints, infrastructure availability, and deployment choices.
When certain locations offer systematic advantages, validators may strategically co-locate to maximize expected rewards, giving rise to location games reminiscent of Hotelling's law~\cite{hotelling1929stability}. Ethereum provides a clear illustration: despite its large validator set, participation is heavily concentrated along the Atlantic corridor~\cite{chainbound_geovalidators}---especially in the United States and Europe---where latency to other validators and infrastructure is most favorable (see~\Cref{fig:centralization101}).

In this paper, we study how Ethereum's protocol design, particularly its block-building architecture, systematically shapes the geographical positioning incentives of validators. Since proposer rewards from block building can constitute a significant component of validator income, particularly for large staking operators that experience proposer opportunities more frequently~\cite{mevboostvalue}, we focus on the two primary block-building paradigms currently used in Ethereum: \emph{local block building}, where validators self-construct blocks and handle dissemination, and \emph{external block building}, where validators outsource block construction and dissemination. We compare the migration incentives induced by these paradigms and evaluate their resilience to geographical centralization across different geographical distributions of information sources and validators, revealing paradigm-dependent and sometimes opposing centralization dynamics, as well as under varying consensus parameters such as attestation thresholds and slot duration.

We emphasize block-building paradigms and their interaction with consensus parameters because Ethereum is actively considering protocol changes that directly affect these dimensions, including an in-protocol block-outsourcing mechanism via \gls{epbs}~\cite{epbs} and halving the slot duration~\cite{EIP7782}. Assessing the resulting geographical positioning incentives under these rules is critical for identifying protocol-level levers that may prevent validator concentration and the erosion of core security and fairness properties.

\parhead{Empirical and theoretical limitations}
While geographical decentralization has been studied through mining power or stake distributions~\cite{cong2021decentralized,grandjean2024ethereum,blockchain2025poolschart} and empirical snapshots of validator locations~\cite{ccaf2025ethereum,ethernodes2025countries,kim2018measuring}, these approaches provide only a partial view. They describe the current distribution but cannot predict counterfactual outcomes or the effects of future protocol changes. Moreover, empirical measurements conflate multiple influences---including infrastructure, protocol-level mechanisms, and heterogeneous validator strategies---making causal attribution difficult. Purely theoretical analyses also face limitations: while they can isolate key mechanisms, capturing the joint effects of heterogeneous strategies, network latencies, and protocol rules in a single tractable model typically requires substantial simplification.

\parhead{A unified analytical and computational framework}
To overcome the limitations of empirical and theoretical approaches, and to accommodate heterogeneous agent behavior and interactions with protocol components under heterogeneous geographic and latency conditions, we propose a unified modeling framework. 
At its core, the framework formalizes validators' location incentives in a structured model that admits both analytical characterization and agent-based simulation~\cite{macal2005tutorial}. 
The analytical characterization yields qualitative predictions about the direction and scaling of validators' location incentives.
We then extend this framework computationally through an agent-based simulation, which enables counterfactual analysis and controlled variation of key factors, allowing us to isolate the effects of block-building paradigms on migration incentives and resulting validator geography. 
Together, the analytical and computational analyses provide a systematic view of how geographic concentration impacts Ethereum's decentralization and liveness.

To the best of our knowledge, this is the first work to provide a unified analytical and simulation-based analysis of how Ethereum's block-building mechanisms, validator and information-source distributions, and consensus parameters jointly shape validators' geographical positioning incentives and the emergence and persistence of spatial concentration.

We make the following contributions:

\begin{itemize}[leftmargin=*, topsep=2pt]
    \item We propose a formal model of validators' geographical positioning incentives under Ethereum's consensus protocol and block-building paradigms. The model captures latency structure, information-source placement, and strategic migration decisions.
    \item We analytically characterize the model under a mean-field approximation, deriving qualitative predictions about the direction and scaling of location-dependent incentives.
    \item We computationally extend the model through an agent-based simulation framework calibrated with real-world latency data, enabling controlled experimentation across geographic, infrastructural, and protocol configurations.
    \item Combining our analytical characterization and computational evaluation, we show that Ethereum's block-building architecture is not geographically neutral. Both paradigms generate location-dependent payoffs and incentives to relocate closer to payoff-relevant parties in order to reduce propagation delays, although through different underlying mechanisms. Asymmetric access to information sources further amplifies geographical centralization. We also demonstrate that consensus parameters, such as attestation thresholds and slot times, modulate latency sensitivity and can amplify these effects.
    \item We discuss the implications of our findings for protocol design and outline potential mitigation directions informed by our analysis.
    \item We release the simulation framework and a public dashboard of our results~\cite{geo2025simulation,geographical-decentralization-simulation_2025} to support reproducibility and future research on geographical decentralization.
    
\end{itemize}

\section{Background}

\parhead{Proof-of-Stake Ethereum}
Ethereum has operated under \gls{pos} since September 2022~\cite{ethereum2022merge}. In \gls{pos}, any participant can become a validator by depositing at least 32 ETH as collateral, which aligns incentives with the network's security~\cite{buterin2020combining}.
Block production under \gls{pos} follows the proposer–attester model. In each 12-second slot, a validator is pseudorandomly selected as the \textit{proposer} to create and broadcast a new block. A committee of validators, also pseudorandomly selected from the active validator set, serves as \textit{attesters} by making attestations on the proposed block.
Attesters verify the proposed block and broadcast attestations reflecting their local fork-choice view of the chain. These attestations contribute weight to fork choice and finality, affecting whether subsequent proposers continue building on the block and whether the chain later finalizes it.
In this paper, we approximate this process with a threshold on attestation weight: a block is treated as canonical once the threshold is exceeded.

\parhead{Maximal Extractable Value}
\gls{mev} refers to the value that privileged entities (e.g., proposers) can capture by inserting, excluding, or reordering transactions within a block~\cite{daian2020flash}.
\gls{mev} poses a threat to decentralization by skewing the reward distribution, which in principle should be homogeneous across validators given equal stake. Since effective extraction requires substantial capital, specialized infrastructure, and sophisticated algorithms, \gls{mev} can foster centralization by concentrating rewards among well-resourced actors~\cite{EthereumFoundationPBS}.

\parhead{Proposer-Builder Separation}  
\gls{pbs} is introduced to enable validators to outsource block building (and \gls{mev} extraction) to specialized entities called \emph{builders}, allowing them to benefit from \gls{mev} rewards regardless of their resources or infrastructure, while reducing centralization pressure. \gls{pbs} is currently implemented out of the Ethereum protocol via MEV-Boost~\cite{flashbots2025mevboost}, where the proposer conducts an auction among builders through a trusted third party (\textit{relay}) to obtain the most valuable block. \gls{pbs} will be integrated at the protocol level (referred to as \gls{epbs}) in Ethereum's upcoming ``Glamsterdam'' upgrade, where relay dependencies are removed, and proposers can access bids from builders directly~\cite{epbs}. Under \gls{pbs}, the MEV of a transaction typically refers to the priority fee plus direct transfer to the builder from order flow providers (e.g., searchers) alongside their transaction ~\cite{oz2024wins,wu2025measuringcexdex}.
A proposer may strategically delay block publishing to gain additional \gls{mev} from transactions arriving later. This behavior is known as timing games~\cite{oz_time_2023,schwarz2023time}.

\parhead{Agent-Based Modeling}
\gls{abm} studies complex systems by specifying individual entities (\emph{agents}) and the rules governing their behaviors~\cite{macal2005tutorial}. It is particularly suited to settings with heterogeneous participants, adaptive decision-making, and interactions that resist closed-form analysis. An \gls{abm} typically consists of agents situated in an \emph{environment}, each following a set of \textit{strategies}. The environment provides context and constraints, while agents respond based on their strategies. By modeling and simulating local interactions, \gls{abm} reveals how individual incentives generate emergent system-level outcomes.

\section{Model}
\label{sec:model}
In this section, we introduce the formal model that governs validators' geographical positioning incentives under two block-building paradigms, local and external block building.
The model specifies geographical and latency assumptions, agent characteristics, environmental components such as the consensus protocol and information sources, and the agents' strategy space. 
It serves as the foundation for the analytical characterization in the next section and the computational evaluation presented later. 
The symbols used throughout the model are summarized in \Cref{tab:symbols}.

\subsection{Base Model}
We first introduce the common model structure shared by both block-building paradigms. The model adopts an agent-based representation to capture the heterogeneous validator strategies and their interactions under the consensus protocol. The system consists of autonomous agents whose incentive-driven behaviors collectively generate global outcomes. The canonical components of the model are as follows.

\parhead{Geographical regions and latency.}
We model geography as a finite set of discrete regions
$\mathcal{R} = \{r_1, r_2, \dots, r_m\}$.
For any two regions $r_j, r_k \in \mathcal{R}$, define the random propagation latency of a message $d(r_j,r_k)$.
We assume $d(r_j,r_k)$ is drawn from a log-normal distribution with parameters known to agents $\ln [d(r_j,r_k)] \sim \mathcal{N}(\mu_{jk}, \sigma^2),$ and the expected latency can be given by $\mathbb{E}[d(r_j,r_k)] = e^{\mu_{jk} + \sigma^2/2}$.
This abstraction simplifies geographical modeling while remaining consistent with empirical network measurement studies: inter-region latency is affected by routing and infrastructure, and therefore does not necessarily increase monotonically with geographical distance~\cite{spring2003causes,subramanian2002geographic}. Moreover, log-normal models have been used as a reasonable approximation for positive, right-skewed network delay variables, including RTT variability in prior work~\cite{fontugne2015empirical}.
The log-normal distribution allows us to capture the heavy-tailed nature of network latency, where rare but high-latency events can affect block propagation and attestation.

\parhead{Agents}
Each validator in the set $\mathcal{V} = \{v_1, \dots, v_P\}$ is modeled as an autonomous agent. 
An agent $v_i$ is characterized by the following attributes:
\begin{itemize}[topsep=2pt, itemsep=1pt, parsep=1pt, leftmargin=*]
    \item \emph{Stake} $s_i \in \mathbb{R}_{\geq 0}$, %
    which determines both the probability of proposer selection and attestation weight. For tractability, we assume that all validators hold equal and constant stakes. %
    \item \emph{Region} $r_i \in \mathcal{R}$, which determines the latency for messages %
    exchanged with other agents.
\end{itemize}

\parhead{Environment} The environment consists of two components: (i) the consensus protocol, which governs time, proposer selection, and attestation; and (ii) information sources, which determine the value of blocks proposed by the validators. We describe each in turn. 

\smallskip \noindent {\it Time and consensus.} 
Drawing from the timing games model in ~\cite{schwarz2023time}, we partition time into slots $n \in \{1,\dots,N\}$ of fixed duration $\Delta > 0$. 
In each slot $n$, a validator is selected as the proposer $p_n$, and a set of attesters 
$\mathcal{A}_n \subseteq \mathcal{V} \setminus \{p_n\}$ is assigned to vote on the proposal\footnote{In Ethereum, the proposer may also be assigned as an attester in the same slot. We exclude the proposer from $\mathcal{A}_n$ as a modeling simplification, since this has a negligible impact on the analysis.}. 
A block becomes canonical if at least a fraction $\gamma \in (0,1]$ of attesters in $\mathcal{A}_n$ 
vote positively before a cutoff time $\tau_{\text{cut}} \in [0, \Delta]$. The timeliness of attester votes depends on both the block's release time and the propagation latencies. 
We abstract away the details of fork-choice rules.

\smallskip \noindent {\it Information Sources: Signal and Supplier.} 
We define an information source $I \in \mathcal{I}$ as an exogenous generator of block value for the proposer. Each source is associated with a geographic region $r(I) \in \mathcal{R}$, which determines its latency to the proposer and, consequently, the freshness of the information it provides. We assume information sources to be fungible, with identical value-generation parameters. To isolate the effects of geography, we vary only their spatial distribution, enabling a controlled comparison between homogeneous and heterogeneous placements and their impact on validator incentives.

We distinguish two types of information sources:  
$\mathcal{I}_{\text{signal}} \subseteq \mathcal{I}$ and $\mathcal{I}_{\text{supplier}} \subseteq \mathcal{I}$. 
A \emph{signal} source contributes only partially to the block value---for example, a centralized exchange providing price information that can be arbitraged on-chain. Proposers aggregate value from multiple such signal sources when constructing a block. 
A \emph{supplier}, by contrast, fully determines the value of the proposed block, as in \gls{pbs}, where an external party delivers a complete block. In this case, the proposer captures value from a single supplier source.

Formally, let $V_I(t)$ denote the value available from source $I$ at elapsed time $t$ within a slot, capturing the temporal evolution of extractable value~\cite{oz_time_2023,schwarz2023time,wahrstatter2023time}. 
Motivated by empirical evidence that the value from a supplier is approximately linear~\cite{thiery2023bbp}, we model $V_I(t)$ as a deterministic linear function \(V_I(t) = a_I t + b_I,\) where $a_I > 0$ denotes the growth rate of extractable value
and $b_I > 0$ is the initial value at the start of the slot, representing transactions accumulated since the previous block.\footnote{The initial value $b_I$ corresponds to the reward a proposer would obtain by releasing the block immediately, i.e., without delaying to exploit timing games.}

\parhead{Strategy Space}
We focus on two strategic choices available to a proposer in a slot:
\begin{itemize} [topsep=2pt, itemsep=1pt, parsep=1pt, leftmargin=*]
    \item \emph{Timing:} The proposer chooses a block release time $\tau \in [0,\tau_{\text{cut}}]$, which affects both attestation success and utility.
    \item \emph{Migration:} Proposers
    may migrate between regions at a cost $c$. 
    We model migration as instantaneous, justified by pre-synchronization of validator nodes across multiple regions.  
\end{itemize}

\subsection{Block-Building Paradigms}
Block building is a critical duty for Ethereum validators and a major source of their rewards. An honest but economically rational validator seeks to optimize along two dimensions: (i) constructing a high-value block by capturing \gls{mev} from transactions, and (ii) ensuring that the block becomes canonical by proposing a valid block on time and reaching a sufficient number of attesters. These objectives are inherently in tension due to timing-game incentives, whereby validators may delay block proposals to increase block value at the risk of missing consensus deadlines~\cite{schwarz2023time, oz_time_2023}.

We focus on the two primary block-building paradigms adopted by Ethereum validators:
\begin{itemize}[leftmargin=*]
    \item \emph{Local Block Building}: The validator collects signals---such as price information and transactions---from a distributed set of information sources (e.g., centralized exchange servers, wallet RPC endpoints) and locally constructs the block. It then signs the block header and propagates the block to the network, aiming to reach sufficient attesters before the consensus deadline.
    \item \emph{External Block Building}: The validator outsources block construction to a third-party supplier, such as a builder in the \gls{pbs} design. The validator blindly signs the provided block header, returns it to the supplier, and relies on the supplier to propagate the block to the network, again aiming to reach sufficient attesters on time.
\end{itemize}

\begin{figure}[hbtp]
    \centering
    \includegraphics[width=0.8\linewidth]{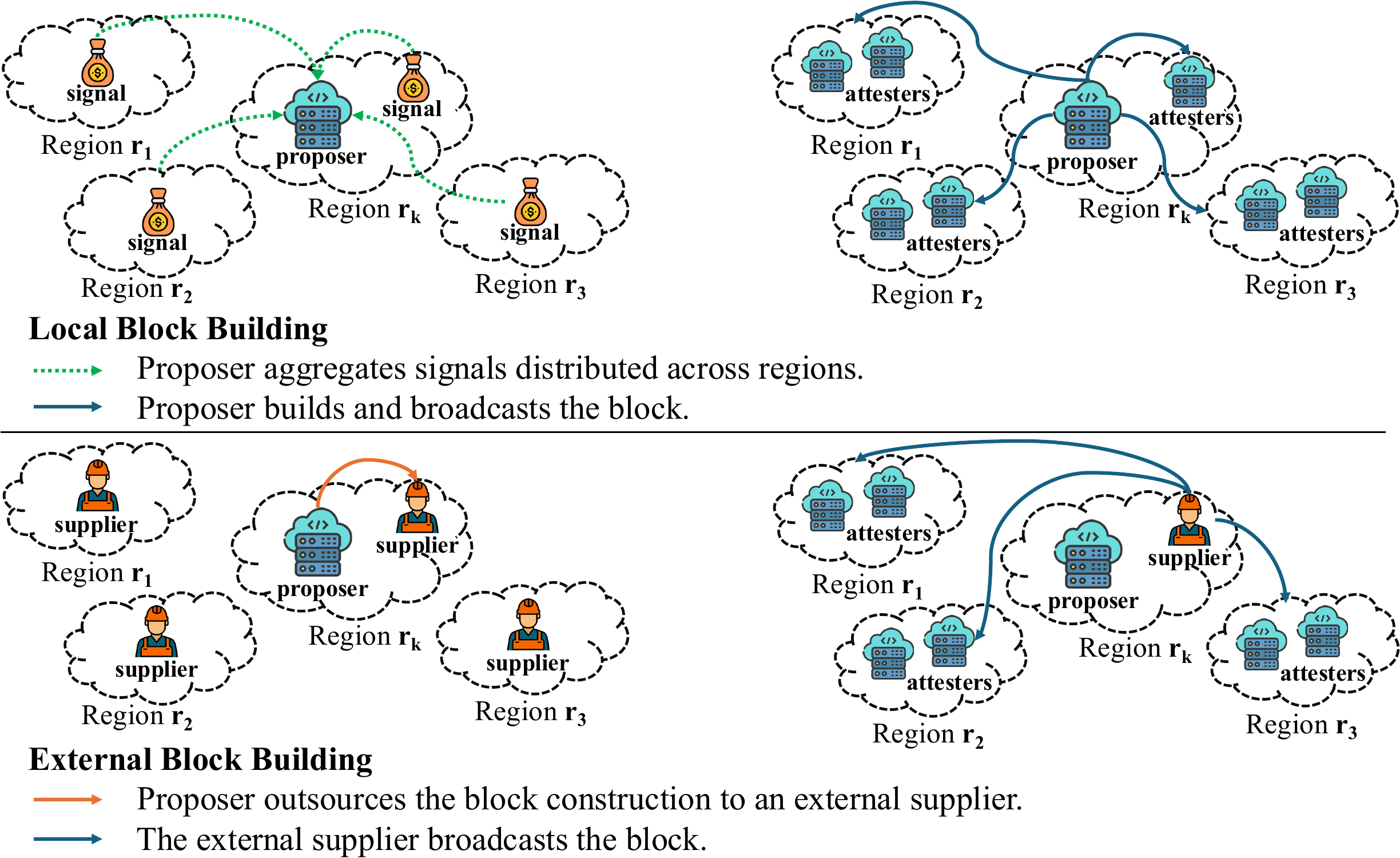}
    \caption{Block production and dissemination under the Local and External block-building paradigms.}
    \Description{}
    \label{fig:block-building-paradigm}
\end{figure}

\Cref{fig:block-building-paradigm} illustrates the operational dynamics of the two block-building paradigms. In practice, external block building dominates, with locally built blocks accounting for less than 10\% of recent Ethereum mainnet blocks~\cite{mevboostpics}.

\subsubsection{Model: Local Block Building}
\label{model:local}
In the local block-building model, a proposer's payoff depends on two latency components: (i) latencies to information sources, which determine how much block value can be incorporated at proposal time, and (ii) latencies to attesters, which determine whether the proposed block is voted canonical after release.

If a proposer $p_n$ located in region $r(p_n)$ releases a block at time $\tau$, it incorporates value from each signal source $I \in \mathcal{I}_{\text{signal}}$ as observed at an earlier time $\tau - d\big(r(p_n), r(I)\big)$, accounting for network latency. The value contributed by source $I$ is therefore
\[
V_I\!\left(\tau - d\big(r(p_n), r(I)\big)\right).
\]

We assume that values from different signal sources are additive, i.e., sources do not generate overlapping transactions. Generalizing to an arbitrary proposer located in region $r \in \mathcal{R}$, the aggregate block value available at release time $\tau$ is given by
\[
V_r(\tau) = \sum_{I \in \mathcal{I}_{\text{signal}}}
V_I\!\left(\tau - d\big(r, r(I)\big)\right).
\]

\noindent \textit{Attestation success.}
Upon release, each attester $a \in \mathcal{A}_n$ located in region $r(a)$ receives the block after a latency $d\!\big(r,r(a)\big)$, drawn from a log-normal with \gls{cdf} $F_{d(r,r(a))}$. An attester $a$ votes on time if $\tau + d\!\big(r,r(a)\big) \le \tau_{\text{cut}}$, which occurs with probability
\[
q_a(\tau; r) = F_{d(r,r(a))}\!\big(\tau_{\text{cut}} - \tau\big)\,\mathbf{1}\{\tau < \tau_{\text{cut}}\}.
\]

For each attester $a \in \mathcal{A}_n$, define an indicator random variable
\[
X_a(\tau;r)=
\begin{cases}
1, & \text{if attester } a \text{ receives and verifies the block in time under release time } \tau \text{ from region } r,\\
0, & \text{otherwise.}
\end{cases}
\]

We model $X_a(\tau;r)$ as an independent Bernoulli random variable with success probability
\[
X_a(\tau;r)\sim \mathrm{Bernoulli}(q_a(\tau;r)),
\]
where $q_a(\tau;r)$ denotes the probability that attester $a$ successfully contributes an attestation before the deadline.
The number of timely attestations is then
\[
S(\tau;r)=\sum_{a\in \mathcal{A}_n} X_a(\tau;r).
\]

Since the success probabilities may differ across attesters, $S(\tau;r)$ follows a Poisson--binomial distribution,  and the probability that the block becomes canonical is
\[
\Pi_r(\tau) = \Pr\!\Big[S(\tau;r)\ge \lceil \gamma\,|\mathcal{A}_n|\rceil\Big].
\]

\noindent \textit{Optimal release time.}
A proposer located in region $r$ aims to release its block as late as possible to maximize $V_r(\tau)$, subject to maintaining a sufficiently high probability of canonicalization. Since $V_r(\tau)$ is modeled as deterministic and the latency distributions are known, the proposer solves, at the start of each slot, the following optimization problem:
\[
\tau_r^\star = \max\{\tau \in [0,\tau_{\text{cut}}] : \Pi_r(\tau) \ge R\},
\]
where $R$ represents the proposer's risk tolerance, capturing its willingness to engage in timing games at the cost of a higher probability that the block fails to reach some attesters. The corresponding expected payoff is
\[
W(r) = \Pi_r(\tau_r^\star)\,V_r(\tau_r^\star).
\]

\begin{remark}
For a large attester committee $|\mathcal{A}_n|$, the randomness of $S(\tau;r)$ concentrates around its mean. By the Law of Large Numbers, defining
\[
\zeta_r(\tau) := \frac{1}{|\mathcal{A}_n|} \sum_{a \in \mathcal{A}_n} q_a(\tau; r),
\]
we obtain the approximation $\Pi_r(\tau) \approx \mathbf{1}\{\zeta_r(\tau) \ge \gamma\}$. Under this approximation, $\tau_r^\star$ corresponds to the critical release time at which $\zeta_r(\tau)$ falls below the attestation threshold $\gamma$.    
\end{remark}

\noindent \textit{Migration decision.}
At the start of each slot, the proposer compares its expected payoff $W(r)$ across all regions $r \in \mathcal{R}$. Since migration incurs a cost $c$, the proposer relocates from its current region $r(p_n)$ only if the marginal benefit of migration exceeds this cost:
\[
\max_{r \in \mathcal{R} \setminus \{r(p_n)\}} W(r) - W\!\big(r(p_n)\big) > c.
\]
In this case, the proposer relocates to
\[
r^\star = \arg\max_{r \in \mathcal{R} \setminus \{r(p_n)\}} W(r),
\]
and releases the block at the corresponding optimal time $\tau_{r^\star}^\star$. Otherwise, it remains in $r(p_n)$ and releases at $\tau_{r(p_n)}^\star$.

This formulation captures proposers' strategic behavior in jointly optimizing release timing and geographical location to maximize expected payoff.

\noindent\textit{Model justification.}
Under local block building, we model proposers as aggregating signals from geographically distributed information sources, with total block value evolving over time. This captures self-building proposer behavior, in which blocks are continuously updated as new order flow arrives. We explicitly account for proposers' latency to attesters to determine how long they can engage in timing games by delaying block release while still satisfying consensus deadlines.

\subsubsection{Model: External Block Building}
In the external block-building model, two aspects differ from the local block-building case: (i) block value is supplied directly by a block supplier rather than aggregated from multiple signal sources, and (ii) block propagation to attesters originates from the selected supplier's region rather than the proposer's. Consequently, the proposer's payoff depends on both its latency to the supplier and the supplier's latency to attesters.

If a proposer $p_n$ located in region $r(p_n)$ considers a block supplier $I \in \mathcal{I}_{\text{supplier}}$ located in region $r(I)$, it commits to (a block provided by) supplier $I$ at time $\tau$. The value captured by the proposer is determined at the commitment time, after which the block content is fixed. The effective block value is therefore $V_I(\tau) = V_I\!\left(\tau - d\!\big(r(p_n), r(I)\big)\right)$.

Note that under the external block-building paradigm, $\tau$ denotes the proposer's commitment time to supplier $I$, rather than a direct release time to attesters. 
By choosing $\tau$, the proposer indirectly determines how late the supplier can begin propagating the block while still satisfying the attestation requirement.

\noindent\textit{Attestation success.}
Once committed by the proposer, a supplier $I$ located in region $r(I)$ broadcasts the block to attesters. Relative to the local block-building model, an additional latency $d\!\big(r(p_n), r(I)\big)$ is incurred between the proposer and the supplier. As a result, the timely-attestation probability becomes supplier-specific:
\[
q_a(\tau; I) 
= F_{d(r(p_n), r(I)) + d(r(I), r(a))}\!\big(\tau_{\text{cut}} - \tau\big)\,\mathbf{1}\{\tau < \tau_{\text{cut}}\}.
\]

The probability that the block becomes canonical is therefore
\[
\Pi_I(\tau) = \Pr\!\Big[S(\tau; I) \ge \lceil \gamma\,|\mathcal{A}_n| \rceil\Big],
\]
where $S(\tau; I)$ is defined analogously to the local block-building case.

\noindent\textit{Optimal release time.}
Conditional on selecting $I$, the proposer chooses a release time
\[
\tau_I^\star = \max\{\tau \in [0,\tau_{\text{cut}}] : \Pi_I(\tau) \ge R\},
\]
and obtains the expected payoff
\[
W(I) = \Pi_I(\tau_I^\star)\,V_I(\tau_I^\star).
\]

\noindent\textit{Migration decision.}
At the start of each slot, the proposer compares the expected payoff obtained by committing to the best available supplier overall---potentially requiring migration---to that obtained from the best supplier accessible without changing location. Let
\[
W^{\text{stay}} = \max_{I \in \mathcal{I}_{\text{supplier}}\!\big(r(p_n)\big)} W(I)
\]
denote the maximum expected payoff achievable without migrating. The proposer migrates if and only if
\[
\max_{I \in \mathcal{I}_{\text{supplier}}} W(I) - W^{\text{stay}} > c.
\]
In this case, it relocates to the region of a supplier
\[
I^\star \in \arg\max_{I \in \mathcal{I}_{\text{supplier}}} W(I),
\]
and releases the block at the corresponding optimal time $\tau_{I^\star}^\star$. Otherwise, the proposer remains in $r(p_n)$ and selects the best locally accessible supplier.

\noindent\textit{Model justification.}
Under external block building, we intentionally abstract away the signal-aggregation process and model each block supplier as delivering a fully constructed block whose value evolves over time. This reflects proposer behavior under \gls{pbs}, where proposers neither observe nor control underlying order-flow sources or builder competition, but instead choose when to select and release a supplied block. 

While, in practice, block suppliers aggregate geographically distributed order flow, these interactions primarily affect the level of block value rather than the propagation dynamics that determine attestation success and canonicalization. Our model therefore isolates the geographical incentives induced by block origination and dissemination---the latter being the focus of this work. Moreover, increasing vertical integration among order-flow providers, builders, and relays further supports modeling suppliers as unified block-providing endpoints from the proposer's perspective~\cite{oz2024wins,yang2025decentralization,wu2025measuringcexdex,dablockauctioninfra}.

\section{Structural Properties of Location Incentives}
\label{sec:theory}

We design four classes of \glspl{exp} based on our model to isolate the impact of block-building paradigms along key dimensions of interest:

\begin{itemize}[leftmargin=*]
    \item \textbf{\gls{exp}~1 (Information-Source Placement Effect):} 
    Assuming a uniformly distributed validator set and holding network, economic, and consensus parameters fixed, we examine how the geographical placement of information sources affects validators' migration incentives under local and external block-building paradigms.
    
    \item \textbf{\gls{exp}~2 (Validator Distribution Effect):} 
    Assuming a uniformly distributed set of information sources and holding network, economic, and consensus parameters fixed, we examine how the initial geographical distribution of validators affects migration incentives under local and external block-building paradigms.
    
    \item \textbf{\gls{exp}~3 (Joint Heterogeneity):} 
    Assuming heterogeneous geographical distributions of both validators and information sources and holding network, economic, and consensus parameters fixed, we examine how joint spatial heterogeneity shapes validators' migration incentives under local and external block-building paradigms.

    \item \textbf{\gls{exp}~4 (Consensus-Parameter Effect):} 
    Assuming uniformly distributed validators and information sources and holding network and economic parameters fixed, we examine how variations in consensus parameters affect validators' migration incentives under local and external block-building paradigms.
\end{itemize}

In this section, we establish analytical properties of validators' location incentives that arise from the model structure. These results characterize qualitative patterns that hold across parameter values and motivate the simulation experiments that follow, which quantify their magnitude and interaction effects.

A key point is that the main directional result in this section, \Cref{thm:latency-payoff}, does \emph{not} rely on the specific log-normal delay family used in the simulations, nor on linear value growth. It uses only monotone value accrual and first-order stochastic comparisons of delays. Parametric assumptions are introduced only later, when we derive sharper quantitative comparisons such as payoff scaling.

\Cref{sec:model} defines the \emph{realized} model at the level of slot-by-slot delays, realized block values, and canonicalization outcomes. In contrast, this section studies the corresponding \emph{expected-payoff counterpart}, which is the appropriate object for the mean-field comparative statics below. Because the model treats propagation delays as random variables, we formulate the value channel using expected source value. For a proposer in region $r$, define
\[
\bar V_r(\tau)
:=
\sum_{I\in \mathcal{I}_{\mathrm{signal}}}
\mathbb{E}\!\left[
V_I\!\left(\tau-d(r,r(I))\right)
\right],
\qquad
\bar V_I(\tau;r)
:=
\mathbb{E}\!\left[
V_I\!\left(\tau-d(r,r(I))\right)
\right].
\]

No specific functional form for $V_I$ is needed for the monotonicity results below. When we later specialize to $V_I(t)=a_I t+b_I$, we do so only to obtain closed-form expressions for the magnitude of latency advantages.

Throughout this section, we work under the large-committee approximation introduced in \Cref{model:local}. We treat the attester distribution as fixed when analyzing the proposer's location choice. This \emph{mean-field assumption} is justified when the validator set $|\mathcal{V}|$ is large: a single validator's relocation negligibly perturbs the attester distribution, so each validator effectively faces an independent optimization problem. Accordingly, the results below should be read as properties of the \emph{mean-field best-response problem}. The endogenous feedback created by simultaneous migration is studied in \Cref{results} via simulations. 

For the external block-building results, we additionally assume the additive end-to-end delay structure and independence between the proposer--supplier and supplier--attester delay components, so that first-order stochastic improvements in the proposer--supplier hop are preserved along the end-to-end proposer--supplier--attester path.

\subsection{Equilibrium Existence}
We first establish that the region selection game admits a pure-strategy equilibrium. 

\begin{definition}[Mean-field region selection game]
Fix an aggregate attester distribution $\lambda$ over regions. The \emph{mean-field region-selection game} is a game in which each validator $v_i\in \mathcal{V}$ chooses a region $r_i\in \mathcal{R}$ and receives payoff $W_i(r_i;\lambda)$, equal to the optimal expected proposer payoff induced by choosing $r_i$, given the aggregate attester distribution $\lambda$.
\end{definition}

\begin{proposition}[Existence of pure Nash Equilibrium]
Under either block-building paradigm, the mean-field region-selection game admits a pure Nash equilibrium.
\end{proposition}

\begin{proof}
Under the mean-field approximation, each validator treats the aggregate attester distribution as exogenous. 
Hence the payoff
to validator $i$ depends only on its own chosen region. Since $\mathcal{R}$ is finite, the best-response
set
\[
\arg\max_{r\in \mathcal{R}} W_i(r;\lambda)
\]
is non-empty for every validator $i$. Any profile in which each validator chooses one of its best-response regions is, therefore, a pure Nash equilibrium.
\end{proof}

Beyond the mean-field regime, the game no longer decouples. In particular, under local block building, validator migration changes the attester distribution itself, which can reinforce co-location incentives. We therefore use the theory below to derive one-agent comparative statics, and use the simulations to quantify the endogenous amplification generated by simultaneous migration.

When validators are already concentrated, the marginal benefit of migration diminishes: relocating provides little additional improvement in attester proximity or block propagation. Therefore, migration incentives weaken as concentration increases, leading to rapid convergence under both paradigms (cf. \Cref{sec:se2}).

\subsection{Latency-Payoff Monotonicity}

We next derive how proposer payoffs respond to changes in propagation delays. These results formalize the intuition that latency advantages create systematic migration incentives, and crucially, that the two paradigms transmit latency advantages through different channels. We provide the proofs of our theoretical predictions in \Cref{sec:proofs}.

Write $X\preceq_{\mathrm{st}} Y$ for first-order stochastic dominance, i.e., $F_X(t)\ge F_Y(t)$ for all $t$.

Under local block building, define
\[
\tau_r^\star
:=
\sup\{\tau\in[0,\tau_{\mathrm{cut}}]:\zeta_r(\tau)\ge \gamma\},
\qquad
W_L(r):=\bar V_r(\tau_r^\star).
\]

Under external block building, let
\[
\zeta_I(\tau;r)
:=
\frac{1}{|\mathcal{A}_n|}
\sum_{a\in \mathcal{A}_n}
q_a(\tau;I,r)
\]
denote the expected fraction of timely attestations when a proposer in region $r$ commits to
supplier $I$ at time $\tau$, and define
\[
\tau_I^\star(r)
:=
\sup\{\tau\in[0,\tau_{\mathrm{cut}}]:\zeta_I(\tau;r)\ge \gamma\},
\quad
W_E(I;r):=\bar V_I(\tau_I^\star(r);r),
\quad
W_E(r):=\max_{I\in \mathcal{I}_{\mathrm{supplier}}} W_E(I;r).
\]

\begin{lemma}[Monotonicity of optimal release time]
\label{lem:tau-monotone}
If $\zeta_1(\tau)\ge \zeta_2(\tau)$ for all $\tau\in[0,\tau_{\mathrm{cut}}]$, then
\[
\sup\{\tau:\zeta_1(\tau)\ge \gamma\}\ge \sup\{\tau:\zeta_2(\tau)\ge \gamma\}.
\]
In particular, if $\zeta_{r_1}(\tau)\ge \zeta_{r_2}(\tau)$ for all $\tau$, then $\tau_{r_1}^\star\ge \tau_{r_2}^\star$, and for any fixed supplier $I$, if $\zeta_I(\tau;r_1)\ge \zeta_I(\tau;r_2)$ for all $\tau$, then $\tau_I^\star(r_1)\ge \tau_I^\star(r_2)$.
\end{lemma}

\begin{lemma}[Monotonicity of payoff in release time]
\label{lem:payoff-monotone}
Assume $\bar V_r(\tau)$ and, for each fixed supplier $I$, $\bar V_I(\tau;r)$ are non-decreasing in $\tau$. Then under local block building, $W_L(r)=\bar V_r(\tau_r^\star)$ is non-decreasing in $\tau_r^\star$, and under external block building, $W_E(I;r)=\bar V_I(\tau_I^\star(r);r)$ is non-decreasing in $\tau_I^\star(r)$.
\end{lemma}

\begin{remark}
If the effective value function is strictly increasing at the optimum, the monotonicity is strict. In particular, this holds under the linear specification $V_I(t)=a_I t+b_I$ with $a_I>0$.
\end{remark}

Combining \Cref{lem:tau-monotone,lem:payoff-monotone} yields the central comparative static.

\begin{theorem}[Latency-payoff monotonicity]
\label{thm:latency-payoff}
Assume each source value function $V_I$ is non-decreasing in its argument. Under either block-building paradigm, reducing propagation delays to payoff-relevant counterparties weakly increases the optimal release time and weakly increases expected payoff. Formally:
\leavevmode
\begin{enumerate}[label=(\alph*)]
    \item \textbf{Local block building.} 
    If $d(r_1,r(I)) \preceq_{\mathrm{st}} d(r_2,r(I))$ for every $I\in \mathcal{I}_{\mathrm{signal}}$, and $d(r_1,r(a)) \preceq_{\mathrm{st}} d(r_2,r(a))$ for every $a\in \mathcal{A}_n$, then $\tau_{r_1}^\star \ge \tau_{r_2}^\star$, and $W_L(r_1)\ge W_L(r_2).$
    
    \item \textbf{External block building.} 
    If $d(r_1,r(I)) \preceq_{\mathrm{st}} d(r_2,r(I))$ for every $I\in \mathcal{I}_{\mathrm{supplier}}$, then, for every supplier $I$, $\tau_I^\star(r_1)\ge \tau_I^\star(r_2)$ and $W_E(I;r_1)\ge W_E(I;r_2)$, and therefore $W_E(r_1)\ge W_E(r_2)$.
\end{enumerate}
\end{theorem}

If, in addition, the effective value functions are strictly increasing at the optimum, then the inequalities in \Cref{thm:latency-payoff} are strict whenever the delay improvement either strictly shifts the feasible frontier or strictly improves the value term at the common optimum. In part (b), strictness of the maximized payoff additionally requires that a strictly improved supplier either already belongs to $\arg\max_I W_E(I;r_2)$ or becomes optimal under $r_1$. Thus, unlike local building, a strict improvement for a dominated supplier need not produce a strict increase in the maximized external-building payoff.

The theorem above identifies the direction of the incentive effect. 
We now specialize to the linear value specification $V_I(t)=a_I t+b_I$. This additional structure is not needed for the directional monotonicity result above, but it allows closed-form comparisons of the magnitude of latency advantages across paradigms.

\begin{remark}[Attester-facing release time]
\label{rem:phi}
For intuition, let $\phi$ denote the time at which block propagation toward attesters begins. Under local block building, the proposer releases directly to attesters, so $\phi=\tau$. Under external block building, if a proposer in region $r$ commits to supplier $I$ at time $\tau$, then propagation begins from the supplier after the proposer--supplier hop, so at the realized delay
level $\phi=\tau+d(r,r(I)).$
Equivalently,
\[
\tau=\phi-d(r,r(I)),
\qquad
V_I\!\left(\tau-d(r,r(I))\right)=V_I\!\left(\phi-2d(r,r(I))\right).
\]
Thus, information source latency enters the value term once under local block building, whereas proposer--supplier latency enters twice under external block building : once through delayed observation of supplier value, and once through the need to commit early enough for the same attester-facing release time $\phi$.
\end{remark}

\begin{proposition}[Signal-source scaling under local block building]
\label{prop:signal-scaling}
Under local block building, the marginal payoff advantage of a latency reduction scales with the number of signal sources. Assume $V_I(t)=a_I t+b_I$. Consider two regions $r_1,r_2$ such that their attester-delay profiles are identical and
\[
\mathbb{E}[d(r_1,r(I))]
=
\mathbb{E}[d(r_2,r(I))]-\delta
\qquad
\text{for all } I\in \mathcal{I}_{\mathrm{signal}},
\]
for some $\delta>0$. Then
\[
\tau_{r_1}^\star=\tau_{r_2}^\star
\qquad\text{and}\qquad
W_L(r_1)-W_L(r_2)=\delta \sum_{I\in \mathcal{I}_{\mathrm{signal}}} a_I.
\]
In particular, if all sources have the same slope $a_I=a$, then
\[
W_L(r_1)-W_L(r_2)=a\,\delta\,|\mathcal{I}_{\mathrm{signal}}|,
\]
which means the local-building advantage scales linearly with the number of signal sources.
\end{proposition}

\begin{proposition}[Single-supplier selection under external block building]
\label{prop:supplier-selection}
Under external block building, the payoff advantage of a latency reduction is determined by the proposer-to-best-supplier delay and does not scale with the number of available suppliers. Assume $V_I(t)=a_I t+b_I$. Suppose that for every supplier $I\in \mathcal{I}_{\mathrm{supplier}}$, moving from $r_2$ to $r_1$ induces a uniform left shift by $\delta>0$ in the proposer--supplier
delay,
\[
F_{d(r_1,r(I))}(t)=F_{d(r_2,r(I))}(t+\delta)
\quad \text{for all } t,
\]
and the corresponding end-to-end proposer--supplier--attester delays shift by the same $\delta$.
Then, for each supplier $I$,
\[
a_I\delta
\;\leq\;
W_E(I;r_1)-W_E(I;r_2)
\;\leq\;
2a_I\delta.
\]
If $\tau_I^\star(r_2)+\delta \leq \tau_{\mathrm{cut}}$, then the upper bound is attained:
\[
W_E(I;r_1)-W_E(I;r_2)=2a_I\delta.
\]
Consequently, 
\[
W_E(r_1)-W_E(r_2)
\;\leq\;
2\delta \max_{I\in \mathcal{I}_{\mathrm{supplier}}} a_I.
\]
Hence, the gain depends only on the single selected supplier and does not scale with $|\mathcal{I}_{\mathrm{supplier}}|$.
\end{proposition}

\begin{remark}
\Cref{prop:signal-scaling,prop:supplier-selection} together predict that local block building exhibits stronger migration incentives than external block building when information sources are numerous: under local building, latency advantages add across all signal sources; under external building, they are bottlenecked by the selected supplier. This is confirmed in the baseline simulation (\Cref{sec:baseline}, \Cref{fig:baseline}).
\end{remark}

\begin{proposition}[Opposing placement sensitivity]
\label{prop:opposing-placement}

The two paradigms exhibit opposing sensitivities to information-source placement. Fix two regions $r_s,r' \in \mathcal{R}$.

\begin{enumerate}[label=(\alph*)]
    \item \textbf{Local block building.}
    Assume $M$ identical signal sources with common slope $a>0$. Suppose $r_s$ is at least as well connected to attesters as $r'$, and
    \[
    \mathbb{E}[d(r',r_s)]-\mathbb{E}[d(r_s,r_s)]
    \;\geq\;
    \mathbb{E}[d(r',x)]-\mathbb{E}[d(r_s,x)]
    \qquad
    \text{for all } x\in \mathcal{R}.
    \]
    Then the payoff gap $W_L(r_s)-W_L(r')$ is maximized when all $M$ signal sources are located in $r_s$.
    
    \item \textbf{External block building.}
    Fix a supplier $I^\star$ located in region $r_s$. Suppose that moving the proposer from $r'$ to $r_s$ induces a uniform reduction $\delta_{r',r_s}>0$ in the proposer--supplier delay, and that the corresponding end-to-end proposer--supplier--attester delays decrease by the same amount. Then the payoff gain from moving the proposer from $r'$ to $r_s$ while continuing to use supplier $I^\star$ is non-decreasing in $\delta_{r',r_s}$. Hence, a supplier located in a more remote region induces stronger migration incentives toward that region.
    
\end{enumerate}
\end{proposition}

\begin{remark}
\Cref{prop:opposing-placement} explains the opposing centralization patterns observed in \Cref{sec:se1}, \Cref{fig:se1}. Local block building rewards joint proximity to signal sources and attesters, so concentrating signals in a well-connected hub amplifies migration toward that hub. External block building instead rewards elimination of the proposer--supplier bottleneck, so co-location incentives are strongest when the relevant supplier is poorly connected to the rest of the network.
\end{remark}

\subsection{Consensus Parameter Interactions}

We now discuss how consensus parameters modulate the location incentives. Unlike the structural results above, the following statements require additional assumptions and are best interpreted as qualitative predictions to be tested in simulation.

\parhead{Attestation threshold~$\gamma$.}
Higher~$\gamma$ forces earlier release, but the effect on cross-region payoff gaps differs by paradigm. Under local block building, if a well-connected region enjoys its largest timing advantage near~$\tau_{\mathrm{cut}}$, raising $\gamma$ moves the optimum into an earlier regime where cross-region differences in feasible release times are smaller, which tends to dampen migration incentives. However, under external block building, higher~$\gamma$ raises the time the supplier needs to reach a $\gamma$-fraction of attesters, making each millisecond of proposer--supplier latency proportionally more costly and amplifying co-location incentives.

\parhead{Slot duration~$\Delta$.}
Shortening the slot affects the relative weight of fixed propagation delays.

\begin{proposition}[Slot duration and reward disparity]
\label{prop:slot-duration}
Assume all signal sources, and all suppliers, share a common growth rate $a>0$. Suppose that $\tau_{\mathrm{cut}}$ is reduced to $\tau_{\mathrm{cut}}'<\tau_{\mathrm{cut}}$, all optimal release times remain interior, and under external block building, the maximizing supplier for each region does not change. Then, under both paradigms, every region's payoff decreases by the same constant, so pairwise payoff differences are unchanged while the mean payoff decreases.
\end{proposition}

\Cref{{prop:slot-duration}} shows why shorter slots can increase normalized inter-regional reward disparity even when the ranking of regions and the aggregate degree of geographical centralization remain essentially unchanged (cf. \Cref{sec:4.2}).

\Cref{tab:predictions} summarizes the theoretical predictions and their corresponding simulation experiments.

\begin{table}[t]
\centering
\caption{Summary of theoretical predictions and corresponding simulation experiments.}
\label{tab:predictions}
\resizebox{\textwidth}{!}{%
\begin{tabular}{l  l  l  l}
\toprule
\textbf{Prediction} & \textbf{Local Block Building} & \textbf{External Block Building} & \textbf{Tested in} \\
\midrule
Centralization direction (\Cref{thm:latency-payoff}) &
Toward low-latency hubs &
Toward supplier locations &
\Cref{sec:baseline} - \ref{sec:se3} \\
\addlinespace
Migration magnitude (\Cref{prop:signal-scaling} \& \ref{prop:supplier-selection}) &
Scales with $|\mathcal{I}_{\mathrm{signal}}|$ &
Bounded by a single supplier &
\Cref{sec:baseline}, \ref{sec:se1} \\
\addlinespace
Source placement (\Cref{prop:opposing-placement}) &
Low-latency placement amplifies &
High-latency placement amplifies &
\Cref{sec:se1}\\
\addlinespace
Higher~$\gamma$ (qualitative prediction)&
Dampens migration &
Amplifies migration &
\Cref{sec:4.1} \\
\addlinespace
Shorter~$\Delta$ (\Cref{prop:slot-duration}) &
Increases normalized reward disparity &
Increases normalized reward disparity &
\Cref{sec:4.2} \\
\bottomrule
\end{tabular}
}
\end{table}

\section{Simulation}
\label{results}

In this section, we use simulation to quantify the magnitude of the location incentives identified in Section~\ref{sec:theory} and to measure the resulting degree of geographical decentralization under each \gls{exp}.
Unless stated otherwise, we use a baseline configuration in which both validators and information sources are uniformly distributed across regions.

\subsection{Evaluation Metrics}
We first introduce metrics that quantify geographical centralization in our simulations.
To characterize how stake is distributed across geographical units (countries/regions; hereafter ``regions''), we use the Gini coefficient (inequality, denoted $\mathrm{Gini}_{\mathrm{g}}$) and the Herfindahl–Hirschman Index (concentration, denoted $\mathrm{HHI}_{\mathrm{g}}$).
To capture variation in proposer incentives across regions, we define the geographical payoff coefficient of variation, $\mathrm{CV}_{\mathrm{g}}$, which measures disparities in the best proposer payoffs across regions within a slot.
Finally, to quantify how geography impacts consensus liveness, we define the geographical liveness coefficient, $\mathrm{LC}_{\mathrm{g}}$, which reflects the minimum number of regions whose failure can break Ethereum's liveness.
Together, these metrics capture stake inequality and concentration, incentive disparities, and resilience to correlated regional failures.

Formally, let \(\mathcal{R}\) denote the set of regions. For each \(r \in \mathcal{R}\), let \(s_r\) be the total effective balance of validators located in \(r\), and let \(S = \sum_{r \in \mathcal{R}} s_r\). Define normalized shares
\[
w_r = \frac{s_r}{S}, \qquad w_r \ge 0, \quad \sum_{r \in \mathcal{R}} w_r = 1.
\]

\noindent\textit{Geographical Gini.}
The geographical Gini coefficient is
\[
\mathrm{Gini}_{\mathrm{g}}
= \frac{1}{2\,|\mathcal{R}|}\sum_{r \in \mathcal{R}} \sum_{r' \in \mathcal{R}} \big| w_r - w_{r'} \big|.
\]
It equals \(0\) under perfectly even stake distribution and increases with inequality (approaching \(1 - 1/|\mathcal{R}|\) in the extreme of all stakes in one region).

\smallskip \noindent {\it Geographical HHI.}
The geographical HHI is
\[
\mathrm{HHI}_{\mathrm{g}} = \sum_{r \in \mathcal{R}} w_r^{\,2},
\]
with \(\mathrm{HHI}_{\mathrm{g}} \in [1/|\mathcal{R}|,\,1]\). Lower values indicate greater dispersion. 

\smallskip \noindent {\it Geographical Payoff Coefficient of Variation (CV).}
Let $\mathcal{W}$ denote the set of best payoffs for the proposer in region $r \in \mathcal{R}$ in one slot, the geographical payoff coefficient of variation is
\[
\mathrm{CV}_{\mathrm{g}}
= \frac{\sigma(\mathcal{W})}{\mu(\mathcal{W})},
\]
where $\mu(\mathcal{W})$ and $\sigma(\mathcal{W})$ are the mean and standard deviation of $\mathcal{W}$, respectively.
It grows as inter-regional disparities increase.

\smallskip\noindent\textit{Liveness coefficient.}
To capture robustness against geographically correlated failures (e.g., natural disasters or country-level regulation), we adapt the Nakamoto coefficient to Ethereum PoS liveness.
Ethereum PoS maintains liveness as long as at least \(2/3\) of the effective stake is online. Sort regions by stake share \(w_r\) in non-increasing order and let \(r_1, r_2, \dots\) denote that ordering. The \emph{liveness coefficient} ($\mathrm{LC}_{\mathrm{g}}$) is the minimal \(k\) such that
\[
\sum_{i=1}^{k} w_{r_i} \;\ge\; \tfrac{1}{3}.
\]
Intuitively, it is the smallest number of (largest) regions whose simultaneous outage suffices to threaten liveness. Larger values indicate greater resilience.

\subsection{Simulation Setup}
\label{sec:setup}
We next specify the setup of our simulation framework. %
\begin{itemize}[topsep=2pt, itemsep=1pt, parsep=1pt, leftmargin=*]
    \item \textbf{Discrete time:} %
    Time is %
    modeled in discrete steps, with each step corresponding to 50ms to approximate the continuous flow of real time.
    \item \textbf{Simultaneity:} The validators %
    act simultaneously at each time step. For example, once the proposer releases a block, attesters in the same region can immediately attest to it.
    \item \textbf{Regions:}  We calibrated the set of regions $\mathcal{R}$ using data from Google Cloud Platform (GCP), which covers 40 regions worldwide. These GCP regions can be further aggregated into seven macro-regions, as detailed in~\Cref{sec:gcp-regions}.
    \item \textbf{Latency:} We calibrated the log-normal distribution $\mu$ such that expected latency equals the empirical mean latency measured between the corresponding regions on Google Cloud~\cite{google2025lookerstudio}. These measurements are based on two-way delays reported by Google Cloud's public inter-region latency data, ensuring that our simulation reflects realistic wide-area network conditions.  
    We use $\sigma = 0.5$ to represent a moderate level of latency variability.
    To assess the sensitivity of our conclusions to this assumption, \Cref{sec:different-sigmas} reports results for alternative values of $\sigma$.
    For each slot, block propagation delays are independently drawn from the corresponding calibrated log-normal distributions.
    \item \textbf{Consensus:} We follow the standard Ethereum consensus parameters~\cite{noauthor_ethereumconsensus-specs_2025} by setting $\Delta = 12$s, $\tau_{\text{cut}} = 4$s, and $\gamma = 2/3$, and assume that all proposers aim to ensure their block becomes canonical with $R=99\%$. Finally, in all cases, we set $|\mathcal{V}|=1,000$ validators and $N=10,000$ slots.\footnote{As shown in \Cref{sec:different-scales}, this configuration preserves similar qualitative trends observed at larger scales while providing a practical balance between computational feasibility and statistical stability.}
    \item \textbf{Information sources:} For signal sources, we set the value parameters $(a_I, b_I) = (0.01,\,0.001)$, and suppliers are parameterized such that their value matches the \emph{aggregate} signal value available to a proposer in the baseline configuration, with $(a_I, b_I) = (0.4,\,0.04)$. That is, a supplier's offered value is scaled to be comparable to the total value that would be obtained by aggregating all signal sources under local block building.
    
    Since our analysis focuses on how information-source placement and propagation latency shape validators' migration incentives, the absolute scale of block value is not central, provided values are matched across paradigms. We therefore adopt fixed reference parameters rather than time-varying values, abstracting away from the value progress and bidding dynamics observed in today's MEV-Boost auctions.
    \item \textbf{Migration cost:}
    We first run a baseline simulation with $c=0$ and, for each of the first \SI{1000}{} slots,\footnote{We focus on the first \SI{1000}{} slots to capture pre-steady-state incentives: with $c=0$ validators quickly relocate toward high-payoff regions, after which marginal benefit collapse toward zero. Using an early window avoids this collapse.} record the marginal benefit
    of migrating to the proposer's best alternative region.
    The resulting distributions of marginal benefits under both block-building paradigms are shown in \Cref{fig:profit-gap-cdf}.
    Following the Pareto principle, we set $c=0.002$, which prevents approximately 80\% of migration events in the baseline simulation under the external block-building paradigm. We present results for alternative migration-cost settings in~\Cref{sec:different-costs}.
    \item \textbf{Random seed:} For each experiment, we repeat the simulation for 20 independent runs using different random seeds. Unless otherwise stated, all reported metrics are presented as mean values across runs. The corresponding 95\% confidence intervals (95\% CI) are reported in \Cref{sec:uncertainty-estimates}.
\end{itemize}

\parhead{Per-slot execution} The simulation framework evolves slot by slot according to the following procedure. At each slot, a proposer and a committee of attesters are sampled under the current validator distribution. The proposer evaluates its expected payoff across candidate regions under the corresponding block-building paradigm. If migration yields a payoff gain exceeding $c$, the proposer relocates before block release; otherwise, it remains in its current region. The proposer then determines the optimal release time, after which block propagation and attestation outcomes are realized using independently sampled latencies.
Metrics are recorded at the end of the slot.

\parhead{Implementation} We implement our simulation framework using Mesa, a Python-based ABM framework~\cite{ter2025mesa}. The framework consists of about 2.5K lines of code. We conducted experiments on a testbed equipped with an Intel Xeon Platinum 8380 CPU (80 cores), and 128 GB of RAM, running 64-bit Ubuntu 22.04 as the operating system.

\subsection{Baseline Configuration and Outcomes}\label{sec:baseline}
We first report outcomes under the baseline configuration, in which both validators and information sources are uniformly distributed across macro-regions and evenly within each macro-region's GCP regions. This avoids bias arising from the uneven global distribution of GCP regions,\footnote{GCP regions are heavily concentrated in Europe, North America, and Asia (see \Cref{sec:gcp-regions}).} ensuring parity in both validator density and aggregate information value across macro-regions.

While MEV-Boost relays (similar to suppliers in our framework) are unevenly distributed across regions today, this asymmetry partly reflects deployment choices specific to the current relay-based architecture rather than protocol constraints.
We therefore use this configuration as a neutral baseline for isolating how protocol design alone shapes validator incentives over a fixed Internet infrastructure. Unless stated otherwise, all subsequent results are interpreted relative to this baseline.

\begin{figure}[t]
    \centering
    \includegraphics[width=0.8\linewidth]{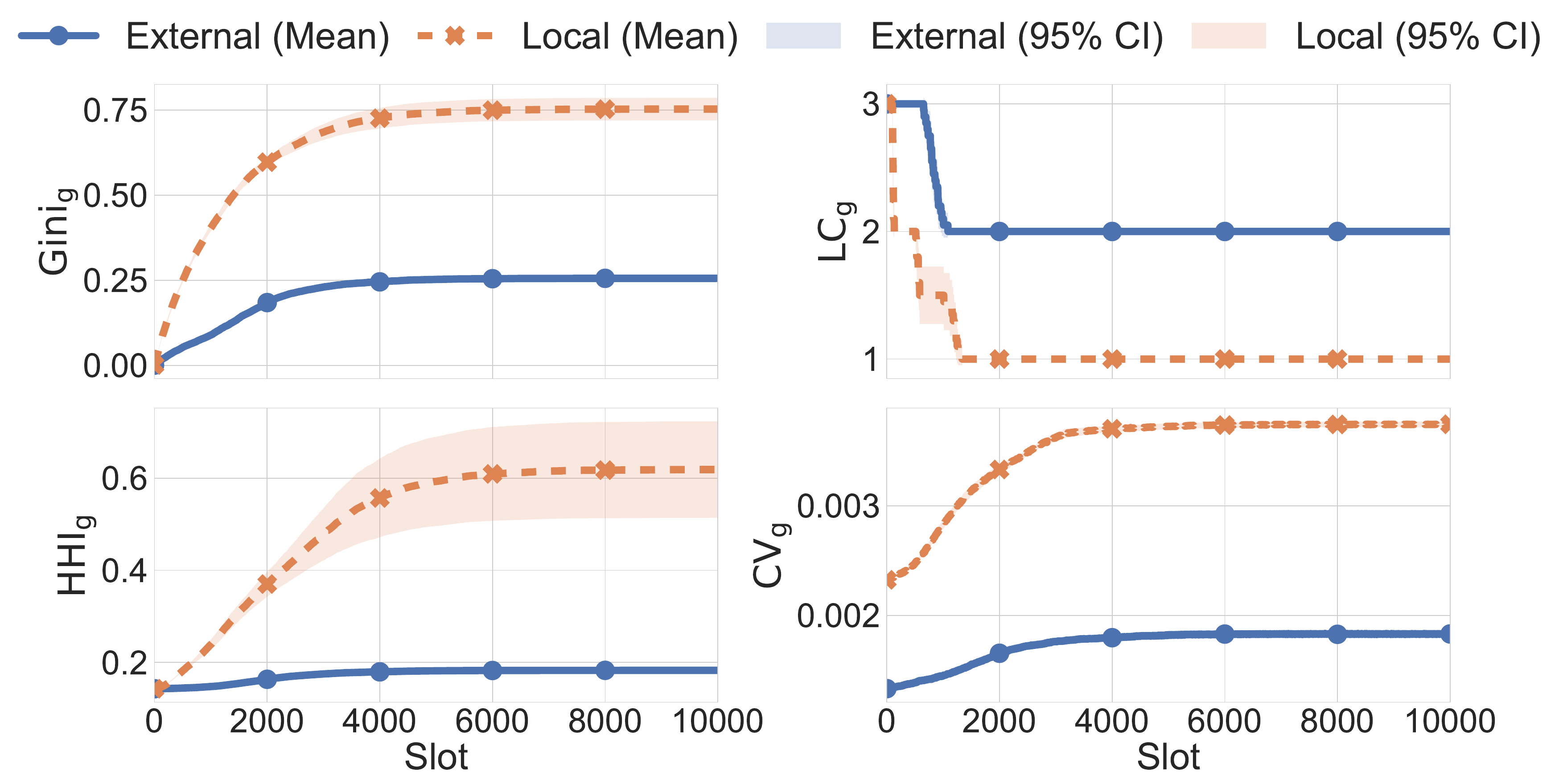}
    \caption{\emph{Baseline configuration (homogeneous validators and information sources).} Evolution of centralization metrics under Local and External block-building paradigms.}
    \Description{}
    \label{fig:baseline}
\end{figure}

\Cref{fig:baseline} shows the evolution of our metrics under the baseline configuration across simulated slots, aggregated over 20 independent runs with different random seeds.
Uncertainty, as reflected by the 95\% CI bands, remains small across all metrics, indicating limited sensitivity to random initialization.
Under both paradigms, the $\mathrm{Gini}_{\mathrm{g}}$ and $\mathrm{HHI}_{\mathrm{g}}$ indices increase over time while the liveness coefficient ($\mathrm{LC}_{\mathrm{g}}$) declines, indicating growing \emph{centralization} even from an initially uniform distribution, consistent with the prediction in~\Cref{thm:latency-payoff}. However, the rate and degree of convergence differ markedly: migration incentives are amplified under the local block-building paradigm, as reflected in faster concentration dynamics and a higher reward-variance coefficient ($\mathrm{CV}_{\mathrm{g}}$).

These differences arise from how migration affects payoffs under each paradigm. Under local block building, proposers are jointly sensitive to the location of signal sources---for value accrual which scales with number of sources (see~\Cref{prop:signal-scaling})---and to the location of other validators---for timely block dissemination. By contrast, when block construction and dissemination are outsourced, block propagation is decoupled from proposer location; the marginal benefit of migration is driven primarily by latency to the selected supplier, which is comparatively less decisive (see~\Cref{prop:supplier-selection}). As a result, migration incentives---and ultimately geographical centralization---are weaker under external block building in the baseline configuration.

In \Cref{sec:convergence_locus}, we present a detailed analysis of the validators' convergence locus.

\subsection{EXP~1: Information-Source Placement Effect}
\label{sec:se1}

Holding validator placement fixed at the baseline configuration, we vary the geographical placement of information sources to examine how validators' migration incentives differ under local and external block-building paradigms. Specifically, we consider two contrasting information-source placements that differ in their average network latency to validators. This design isolates how protocol design interacts with information-source geography to shape migration incentives.

We consider the following two configurations:
\begin{itemize}[leftmargin=*]
    \item \emph{Latency-aligned:} Three information sources are placed in low-latency regions---\texttt{asia-northeast1}, \texttt{europe-west1}, and \texttt{us-east4}---yielding low average latency to validators.
    \item \emph{Latency-misaligned:} Three information sources are placed in high-latency regions---\texttt{africa-south1}, \texttt{australia-southeast1}, and \texttt{southamerica-east1}---resulting in significantly higher average latency.
\end{itemize}

Across both configurations, supplier value parameters and signal-source calibration are kept identical to the baseline, ensuring that observed differences arise solely from geographical placement rather than changes in aggregate information value.

\begin{figure}[t]
    \centering
    \includegraphics[width=0.78\linewidth]{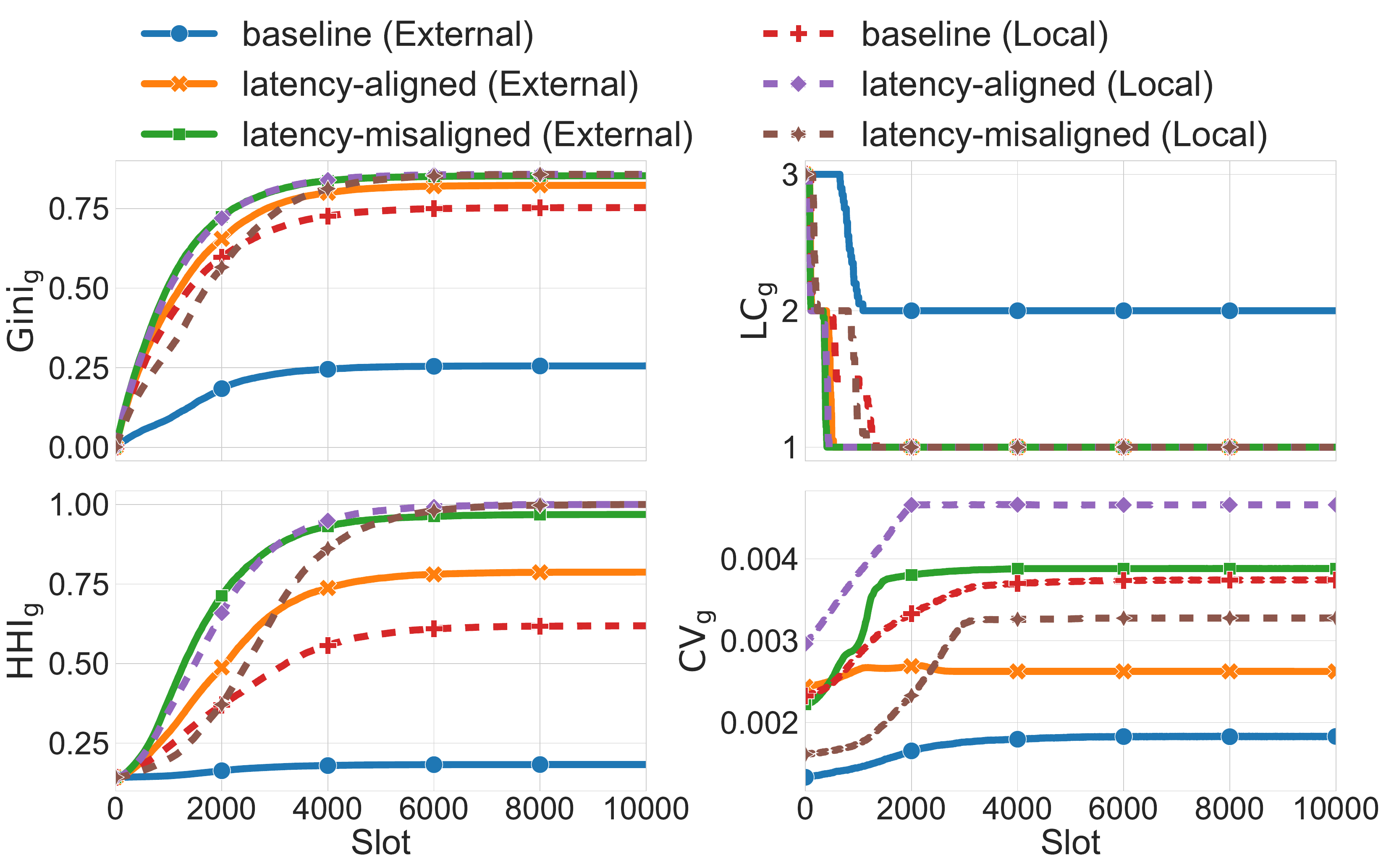}
    \caption{\emph{Information-source placement effect (homogeneous validators).} Evolution of centralization metrics under Local and External block-building paradigms for latency-aligned and latency-misaligned information-source placements.}
    \Description{}
    \label{fig:se1}
\end{figure}

\Cref{fig:se1} shows the evolution of our metrics under latency-aligned and latency-misaligned information-source placements, together with the baseline configuration as a reference.
Given the narrow uncertainty observed in the baseline configuration, subsequent figures report mean trajectories only, while 95\% CI values for final-slot metrics are summarized in \Cref{sec:uncertainty-estimates}.
Under both block-building paradigms, asymmetric information-source placement leads to faster and stronger geographical centralization, a sharper decline in the liveness coefficient, and higher reward variance across regions.\footnote{An exception arises under the local block-building paradigm in the latency-misaligned configuration, which both starts and converges to a lower reward-variance coefficient than the baseline. This behavior likely reflects a trade-off between high-latency regions hosting the information sources, which offer higher marginal value growth over time but suffer from slower propagation to attesters, and low-latency regions, which exhibit the opposite characteristics. The resulting tension leads to a more balanced distribution of rewards across regions relative to the baseline.} This is consistent with the intuition that asymmetries in access to rewards amplify validators' incentives to migrate.

However, the two paradigms exhibit opposite sensitivities to the specific placement of information sources as predicted by~\Cref{prop:opposing-placement}. Under local block building, the latency-aligned configuration centralizes more rapidly, with lower liveness and higher reward variance than the latency-misaligned case. In contrast, under external block building, the latency-misaligned configuration exhibits stronger centralization dynamics. 

The stronger centralization observed under asymmetric information-source placement arises from how migration interacts with source geography under each block-building paradigm. 

Under local block building, locating in a low-latency region simultaneously improves access to signal sources and propagation to attesters, increasing the marginal benefit of migration and making it more likely to exceed the migration cost. Under external block building, by contrast, when suppliers are located in poorly connected regions, non-colocated proposers face large proposer–supplier latencies; migrating closer to the supplier therefore yields a larger payoff increase, more frequently justifying relocation. In this paradigm, the proposer can influence only its latency to the supplier, while proximity to other validators provides no direct benefit, as block dissemination is handled by the supplier.

Overall, these results show that unbalanced information-source placement amplifies migration incentives relative to the balanced baseline. The extent and direction of this amplification depend critically on the adopted block-building paradigm, with different placements either accelerating or dampening geographical centralization.

\subsection{EXP~2: Validator Distribution Effect}
\label{sec:se2}

Holding information-source placement fixed at the baseline configuration, we vary the initial geographical distribution of validators to examine how migration incentives differ under local and external block-building paradigms. Specifically, we initialize validators according to Ethereum's current geographic distribution~\cite{chainbound_geovalidators}, which exhibits substantial concentration in the United States and Europe (see~\Cref{fig:centralization101}). This design isolates how protocol design interacts with validator geography to shape migration incentives.

\begin{figure}[t]
    \centering
    \includegraphics[width=0.8\linewidth]{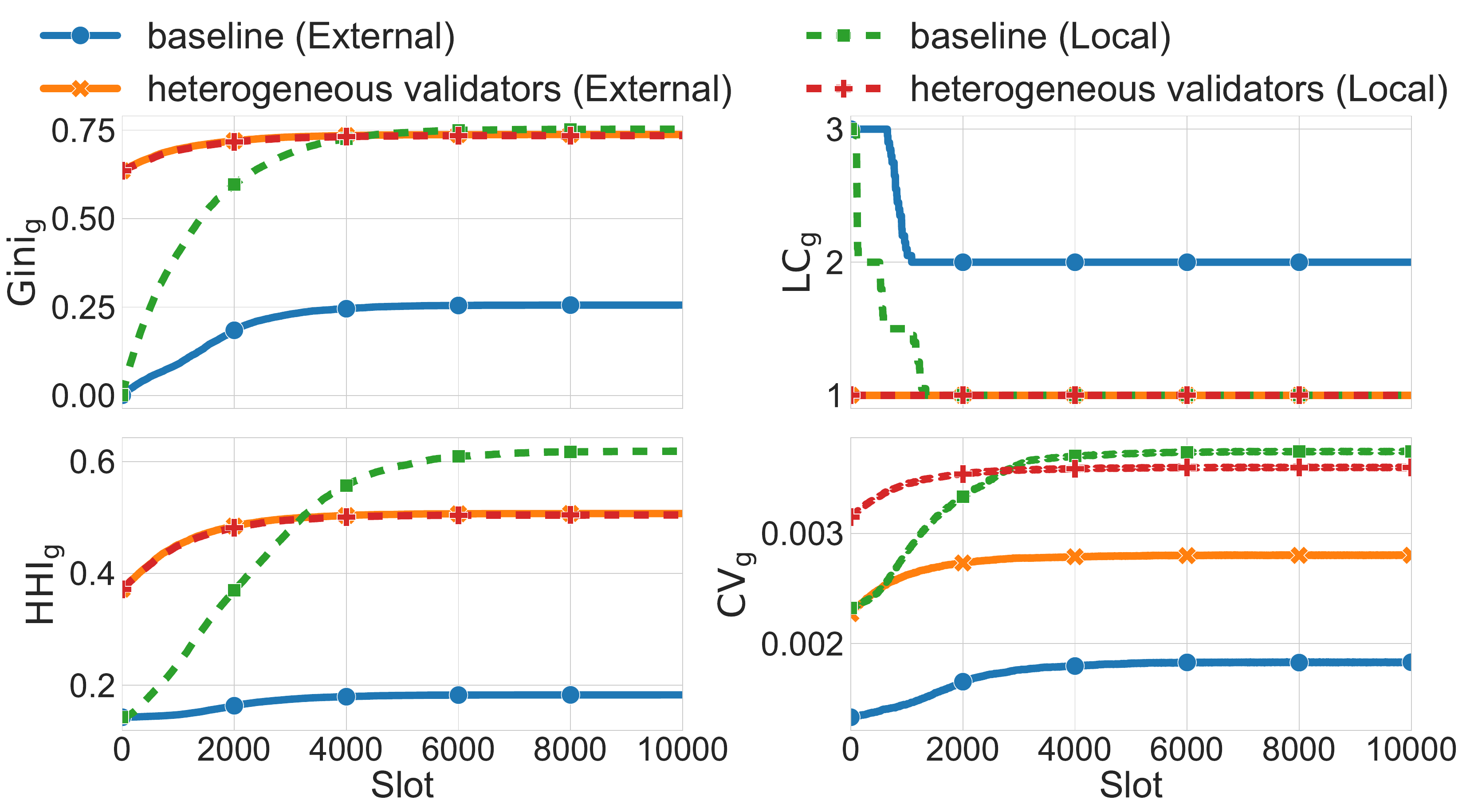}
    \caption{\emph{Validator distribution effect (homogeneous information sources).} Evolution of centralization metrics under Local and External block-building paradigms with a heterogeneous validator distribution.}
    \Description{}
    \label{fig:se2}
\end{figure}

\Cref{fig:se2} shows the evolution of our metrics under a heterogeneous validator distribution, together with the baseline configuration as a reference. Because the initial validator distribution is already geographically concentrated, centralization metrics are elevated and the liveness coefficient is lower at the outset compared to the baseline, which starts from a neutral configuration. Reward variance is likewise higher initially, reflecting pre-existing regional disparities in validator density.

Under both block-building paradigms, this initial concentration leads to rapid convergence toward an equilibrium in which most validators become co-located. Once this state is reached, the marginal benefits of further migration no longer exceed the migration cost. Unlike in the baseline case, the two paradigms exhibit no substantial differences in either the speed or degree of convergence: when incumbent hubs already exist, migration benefits quickly saturate under both paradigms, whereas in the baseline configuration, such hubs emerge endogenously over time and primarily under local block building.

Consistent with earlier results, reward variance remains higher under local block building. However, relative to its baseline, external block building exhibits a stronger amplification of inter-regional reward disparities when validators are already clustered, as propagation advantages translate more directly into payoff asymmetries. By contrast, under local block building, rapid convergence to a dominant hub limits further differentiation, causing reward variance to stabilize once migration saturates.

Overall, these results indicate that when validator geography is already centralized, the currently active block-building paradigms in Ethereum induce similar migration dynamics, with rapid convergence toward co-location and limited scope for further relocation.

\subsection{EXP~3: Joint Heterogeneity}
\label{sec:se3}

We jointly vary the geographical placement of information sources and the initial distribution of validators to examine how migration incentives differ under local and external block-building paradigms when both dimensions are heterogeneous. Specifically, we combine the latency-aligned and latency-misaligned information-source placements from \Cref{sec:se1} with the empirically observed validator distribution from \Cref{sec:se2}. Results are compared against the baseline configuration in which both validators and information sources are uniformly distributed.

This experiment captures interaction effects between validator geography and information-source placement, illustrating how joint spatial heterogeneity shapes migration incentives beyond the effects observed when each dimension is varied in isolation.

\begin{figure}[t]
    \centering
    \includegraphics[width=0.8\linewidth]{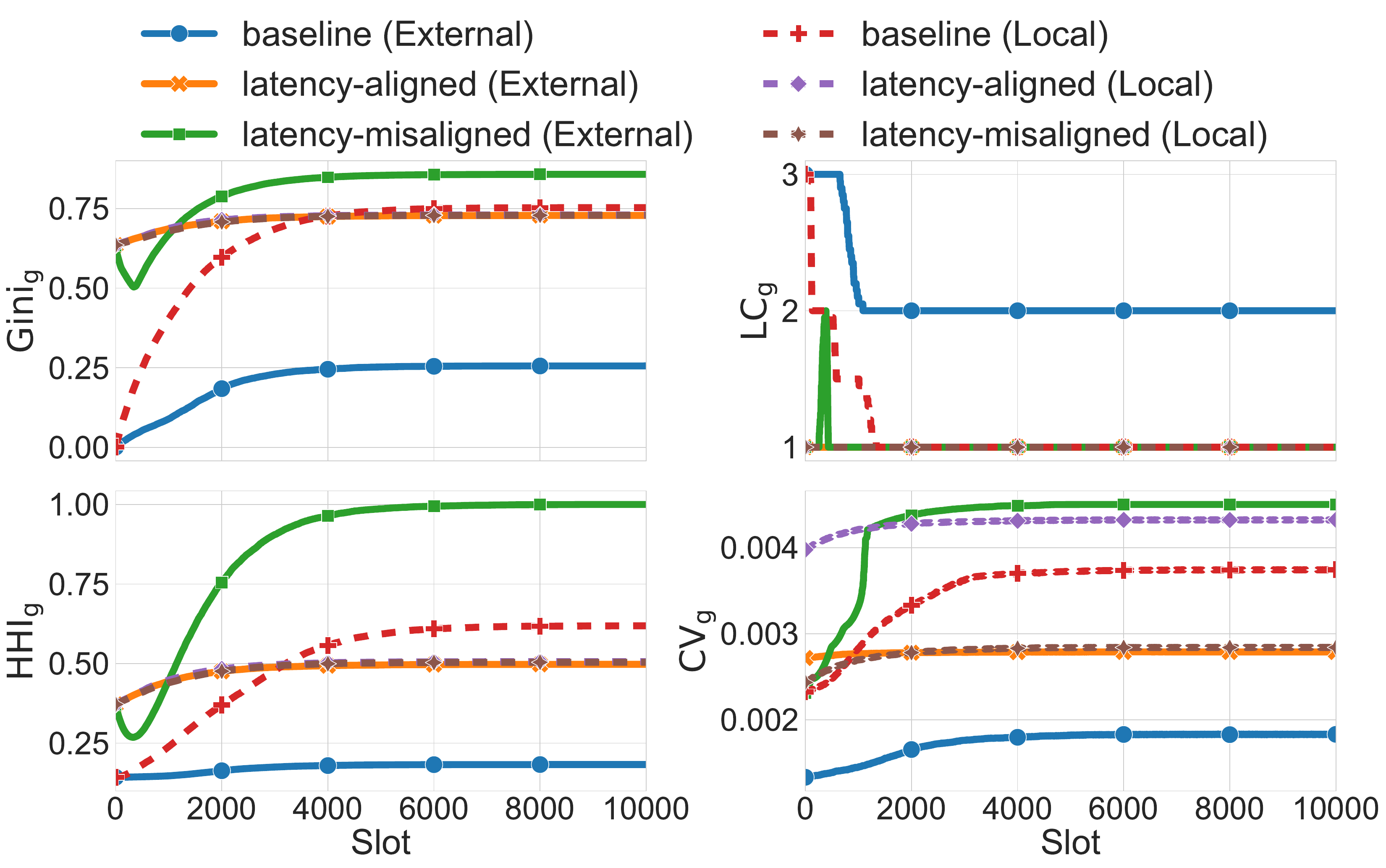}
    \caption{\emph{Joint heterogeneity (validators and information sources).} Evolution of centralization metrics under Local and External block-building paradigms for latency-aligned and latency-misaligned information-source placements with heterogeneous validator distribution.}
    \Description{}
    \label{fig:se3}
\end{figure}

\Cref{fig:se3} shows the evolution of our metrics under latency-aligned and latency-misaligned information-source placements with a heterogeneous validator distribution, together with the baseline configuration as a reference. Overall, the results closely mirror those in \Cref{sec:se1} and \Cref{sec:se2}: as in \Cref{sec:se1}, asymmetric information-source placement leads to stronger geographical centralization, and as in \Cref{sec:se2}, starting from an already concentrated validator distribution results in rapid convergence toward an equilibrium in which the marginal benefits of further migration no longer exceed the migration cost.

The primary deviation arises under the external block-building paradigm when information sources are placed in high-latency regions. As in \Cref{sec:se1}, co-locating with suppliers in poorly connected regions yields a large payoff gain, since proposer–supplier latency is the only delay the proposer can directly influence in this paradigm. In the joint heterogeneity setting, validators are initially concentrated in low-latency regions, reflecting today's Ethereum geography. This produces a transient phase in which some validators migrate away from incumbent hubs, briefly improving geographical decentralization. Over time, because block propagation is handled by the supplier and proximity to other validators provides no direct benefit, incentives to co-locate with suppliers dominate, leading to convergence toward a single supplier region and renewed geographical centralization.
In \Cref{sec:full-hetero-cluster}, we present a detailed analysis of validators' convergence locus under this joint heterogeneity setting.

Overall, these results indicate that under joint spatial heterogeneity of validators and information sources, the system starts from a highly centralized state and rapidly converges to a saturated equilibrium in which further migration yields little benefit. While the strength of these effects depends on the adopted block-building paradigm and the placement of information sources, both settings ultimately exhibit stronger geographical centralization than the baseline configuration, weakening decentralization guarantees.

\subsection{EXP~4: Consensus-Parameter Effect}
\label{sec:se4}

Holding validator and information-source placements fixed at the baseline configuration, we vary consensus parameters to examine how validators' migration incentives differ under local and external block-building paradigms. Specifically, we consider (i) different attestation threshold values and (ii) different slot times.

\subsubsection{Attestation Threshold Effect}
\label{sec:4.1}
We analyze how changing the required attestation threshold $\gamma$---the quorum needed for a block to become canonical---affects migration incentives. We consider $\gamma \in \{\tfrac{1}{3}, \tfrac{1}{2}, \tfrac{2}{3}, \tfrac{4}{5}\}$, reflecting alternative design choices that trade off safety, liveness, and economic penalties, and have been actively discussed in recent consensus design work~\cite{buterin2025threshold}.

\begin{figure}[t]
    \centering
    \includegraphics[width=0.8\linewidth]{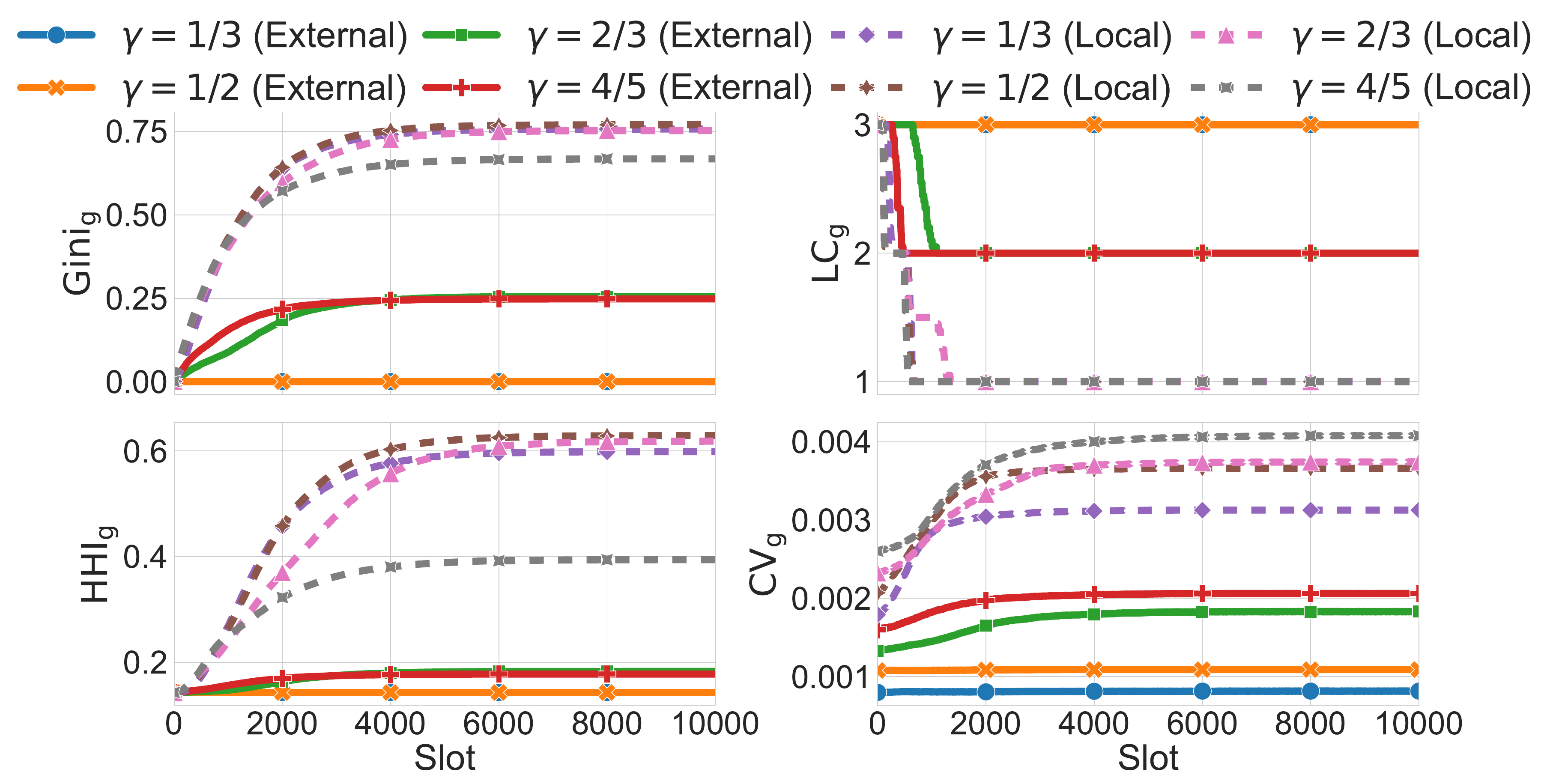}
    \caption{\emph{Attestation-threshold effect (homogeneous validators and information sources).} Evolution of centralization metrics under Local and External block-building paradigms for different attestation thresholds $\gamma \in \{\tfrac{1}{3}, \tfrac{1}{2}, \tfrac{2}{3}, \tfrac{4}{5}\}$.}
    \Description{}
    \label{fig:se4.1}
\end{figure}

\Cref{fig:se4.1} shows the evolution of our metrics under different attestation thresholds, including the baseline value. Across all thresholds, centralization pressure is consistently stronger under local block building, reflected in higher Gini and HHI indices, lower liveness coefficients, and greater reward disparity. However, the impact of varying $\gamma$ differs markedly between the two paradigms. Increasing $\gamma$ tightens the effective timing window for block proposal, as a larger fraction of attesters must be reached before the consensus deadline, thereby increasing the role of latency.

Under external block building, this amplifies migration incentives. Since block dissemination is handled by the supplier, the proposer controls only its latency to the supplier, which matters for both freshness and end-to-end timing, while supplier–attester propagation is exogenous. As the effective timing window shrinks, reducing proposer--supplier latency yields a larger marginal increase in attainable block value, making migration toward the supplier more attractive.

Under local block building, by contrast, higher attestation thresholds dampen migration incentives. Because proposer location affects both value accrual and timely block dissemination, proposers face a sharper trade-off between remaining close enough to attesters to satisfy the quorum and remaining close to multiple information sources to maximize aggregate block value. At lower thresholds, fewer attesters need to be reached, allowing validators to prioritize information-source proximity and migrate toward dominant hubs. Higher thresholds force a more balanced consideration of these two dimensions, reducing the benefit of migrating toward any single region.

Overall, these results show that the impact of the attestation threshold on migration incentives depends critically on the adopted block-building paradigm, as $\gamma$ alters the relative importance of propagation latency and information aggregation in proposers' location decisions.

\subsubsection{Shorter Slot Time Effect}
\label{sec:4.2}
We analyze how changing the consensus slot time $\Delta$ affects validators' migration incentives. Motivated by recent community discussions surrounding \gls{eip}-7782~\cite{EIP7782}, which proposes reducing the slot time from $\Delta=12\,\mathrm{s}$ to $\Delta=6\,\mathrm{s}$ with an earlier attestation deadline of $\tau_{\text{cut}}=3\,\mathrm{s}$, we halve the slot time and shift the attestation deadline accordingly in our simulations. The stated goals of this proposal are increased throughput and improved user experience. Recent empirical measurements~\cite{silve20256sslot} report sub-second block propagation times, suggesting that such shorter slot durations may be feasible in practice.

\begin{figure}[t]
    \centering
    \includegraphics[width=0.8\linewidth]{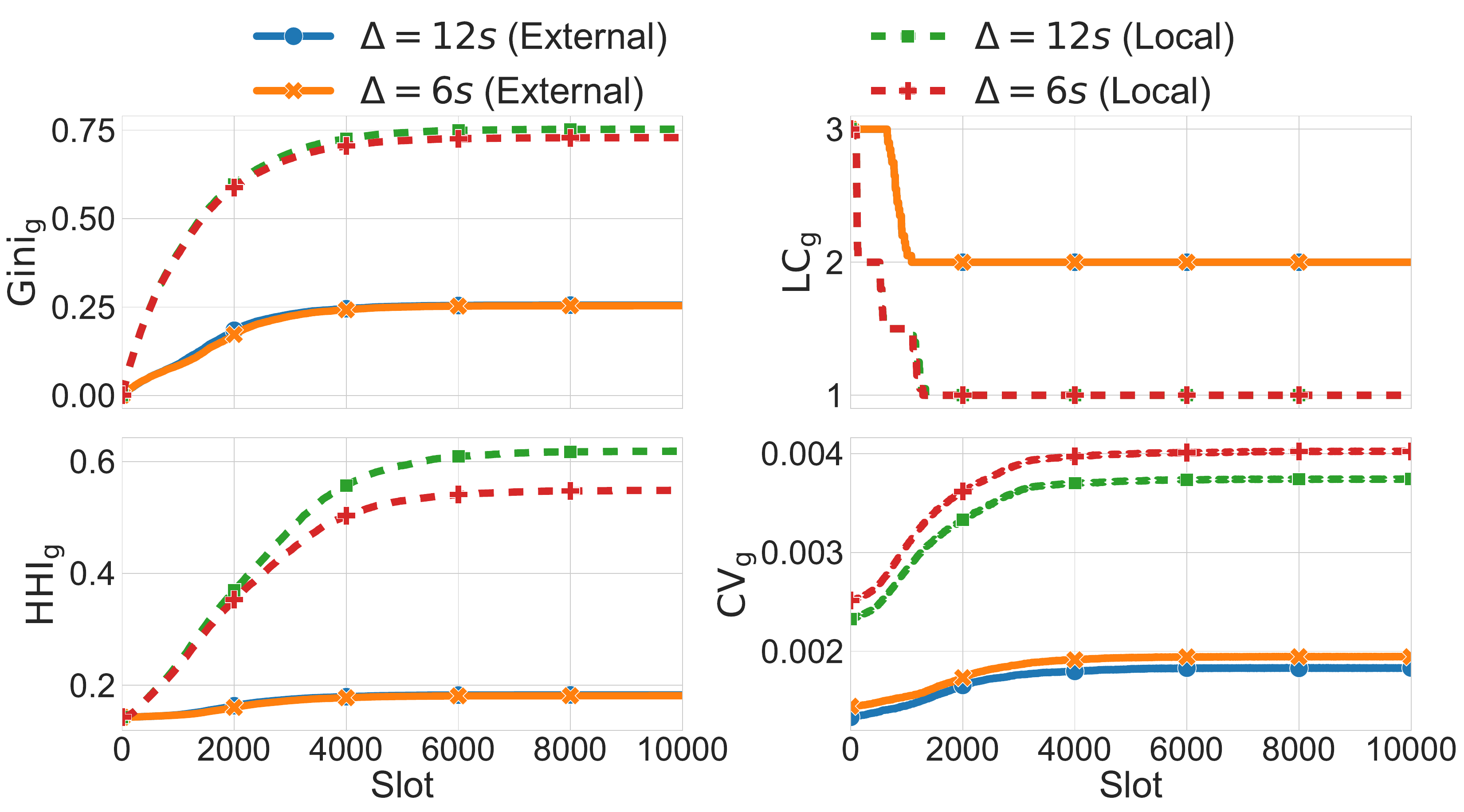}
     \caption{\emph{Shorter slot time effect (homogeneous validators and information sources).} Evolution of centralization metrics under Local and External block-building paradigms with a shorter slot time ($\Delta=6\,\mathrm{s}$) versus the current slot time ($\Delta=12\,\mathrm{s}$).}
     \Description{}
    \label{fig:se4.2}
\end{figure}

\Cref{fig:se4.2} shows the evolution of our metrics under a \SI{6}{\second} slot time, alongside the baseline \SI{12}{\second} configuration. Because observed propagation delays are small relative to the shortened timing window, proposers face essentially the same co-location trade-offs as under $\Delta=12\,\mathrm{s}$. As a result, the trajectories of the centralization indices and the liveness coefficient remain largely unchanged across both block-building paradigms, consistent with empirical observations in~\cite{silve20256sslot}.

We do, however, observe higher reward variance across regions under $\Delta=6\,\mathrm{s}$ for both paradigms, as predicted in~\Cref{prop:slot-duration}. This arises because latency advantages are effectively absolute: when the slot duration is shorter and, as a consequence, the cutoff time is sooner, a fixed latency edge constitutes a larger fraction of the available timing window. Consequently, the same latency differential yields a larger relative payoff advantage, amplifying cross-regional reward disparities.

Overall, these results show that halving the slot time does not materially affect geographical centralization under either block-building paradigm. However, as the relative importance of latency increases, reward disparities across regions become more pronounced, suggesting that beyond a certain threshold, further reductions in slot time may begin to strengthen migration incentives.

\section{Discussion}
In this section, we discuss the implications of our findings, outline how they can inform potential mitigation strategies, and acknowledge the limitations of our framework.

\subsection{Implications}
Our results indicate that Ethereum's block-building architecture is \emph{not} geographically neutral and systematically shapes validators' migration decisions by creating incentives to reduce propagation delays to payoff-relevant parties. Across all experiments, both local and external block-building paradigms induce location-dependent payoffs, albeit through different mechanisms. While externalizing block building via \gls{pbs}/\gls{epbs} decouples block propagation from proposer location, it also creates new incentives to co-locate with suppliers, making supplier geography a central determinant of validator positioning.

More broadly, our findings indicate that information-source placement and supplier location are first-order protocol considerations rather than mere infrastructural artifacts. As shown across heterogeneous source and validator configurations, asymmetries in access to information or block suppliers consistently amplify migration incentives and accelerate geographical concentration, even when validators begin from a balanced distribution.

Finally, consensus parameters act as incentive amplifiers by modulating latency sensitivity. Changes to attestation thresholds and slot times do not uniformly affect centralization outcomes, but they alter which latency components dominate proposers' payoff calculations. As a result, seemingly orthogonal protocol choices---such as quorum requirements or slot duration---can materially influence geographical positioning incentives.

Taken together, these results imply that protocol designers and infrastructure operators must treat geography as an endogenous outcome of protocol design. Ignoring these incentive effects risks undermining decentralization, liveness, and censorship resistance, even in the absence of explicit coordination among validators.

\subsection{Mitigations}
The mechanisms identified in our analysis point to several potential directions for mitigating geographical centralization at both the protocol and infrastructure levels. A key driver of migration incentives in our analysis is the deterministic selection of a single proposer under Ethereum's \gls{pos}, which grants the proposer temporary monopoly power over block release and enables timing games. Our results confirm that this monopoly incentivizes validators to reduce latency in order to extend their effective proposal window, thereby increasing attainable rewards.

One potential mitigation is to weaken the monopoly power of a single proposer through decentralized block-building networks such as BuilderNet~\cite{buildernet} and \gls{mcp} designs~\cite{neuder2024mcp,garimidi_multiple_2025}, in which multiple entities contribute block content within a single slot. However, these protocols may introduce new strategic positioning incentives, as participants seek preferential access to exclusive order flow or content sources, implying a trade-off rather than a definitive remedy.

A complementary mitigation direction operates at the incentive level by reducing the sensitivity of validator rewards to marginal latency advantages. Mechanisms such as \emph{MEV-burn}~\cite{mevburn} dampen the payoff differences arising from timing games, thereby weakening the economic incentive to migrate for latency advantages. While such mechanisms are primarily motivated by economic fairness, our results suggest they may also reduce geographically driven concentration by decoupling reward access from physical proximity.

Finally, our findings indicate that supplier geography~\cite{dageodec} itself becomes a critical determinant of validator positioning under external block building. This suggests that encouraging geographical diversity among builders, relays, or other block suppliers, as well as among major applications that act as signal sources---whether through governance norms, transparency requirements, or explicit protocol incentives---could mitigate the concentration pressures we observe. More broadly, treating geographical decentralization as an explicit protocol objective, rather than an emergent property, may help prevent the long-term erosion of Ethereum's security assumptions.

\subsection{Limitations}

\parhead{Latency data}
Our simulations rely on latency and regional information derived from GCP data. This choice provides a consistent, publicly available, and well-documented dataset, making the analysis transparent and reproducible. Nevertheless, it may introduce bias: other cloud service providers may offer denser regional coverage or slightly lower inter-region latency, potentially affecting validator connectivity and block propagation dynamics, and thus the quantitative outcomes of our simulations.

That said, major providers such as AWS, Azure, and OVH, exhibit broadly similar regional distributions~\cite{AWS2025GlobalInfrastructure,Microsoft2025Azure,OVHcloud2025}, with most regions concentrated in Europe, North America, and Asia. Moreover, since inter-regional latencies are largely determined by underlying fiber-optic infrastructure, using data from alternative providers would likely yield comparable results. As future work, we plan to use datasets from other providers to further validate our findings.

\parhead{Value function}
We model block value as a deterministic function that increases monotonically over time. This abstraction is chosen to isolate proposer timing and migration incentives while ensuring tractable analysis. However, it omits stochastic effects present in practice, such as random transaction arrivals, volatile \gls{mev} opportunities, or builder-specific bidding dynamics. Under the local block-building paradigm, we further assume that signals from different information sources are additive. In practice, such signals may overlap, and their joint contribution to the block value would depend on their interaction structure, which we do not model. Finally, we treat information sources as fungible with identical value parameters, whereas in practice, suppliers and signal sources may differ substantially in the value they provide.

\parhead{Information assumptions}
Our framework assumes that proposers have sufficiently accurate information to evaluate their decisions, including expected propagation delays across regions and expected value differences between locations. This corresponds to a full-information benchmark that allows us to characterize proposers' best-response behaviors in an interpretable manner.

In practice, such information is often only available at coarse granularity. Expected values depend on stochastic market dynamics, and validators may have limited visibility into the geographical distribution of other validators, constraining their ability to form precise expectations about propagation delays. Imperfect information would primarily affect the speed and path of adjustment rather than the existence of incentives themselves. With noisy estimates, proposer behavior can be interpreted as perturbed best responses, leading to slower convergence and occasional mis-relocations, but not eliminating the underlying latency- and value-driven incentives identified in our analysis.

\parhead{Migration} 
We assume that validator migration occurs instantaneously and at a constant cost. This is a best-case abstraction that allows us to focus on long-run location incentives rather than short-run adjustment frictions.
In Ethereum, validators can anticipate proposer duties in advance and pre-synchronize infrastructure across regions, which can make near-instant switchover feasible.
In practice, however, migration may involve not only a one-time switching cost, but also persistent region-specific costs and constraints, such as technical frictions, cloud-provider pricing, regulatory factors, and operational risk. 
Modeling heterogeneous and time-varying migration processes is an important direction for future work.

\section{Related Work}

\parhead{Blockchain decentralization}
Prior work has measured (de)centralization using diverse metrics.
The Nakamoto coefficient was introduced to capture the minimum number of entities needed to compromise a blockchain subsystem~\cite{srinivasan2017quantifying}.
Other measures, including the Gini coefficient~\cite{dorfman1979formula}, Shannon entropy~\cite{lin1991divergence}, and the Herfindahl–Hirschman index~\cite{rhoades1993herfindahl}, have been applied to quantify %
(de)centralization in various contexts, such as mining power~\cite{kwon2019impossibility,lin2021measuring,wu2019information}, staking distribution~\cite{grandjean2024ethereum,kwon2019impossibility}, and block building~\cite{heimbach2023ethereum,oz2024wins,yang2025decentralization}.
Our work examines a different dimension by proposing metrics that capture the geographical distribution of validators.

\parhead{Geographical location of Ethereum validators}
Validators' geo-location has received less attention despite its importance for latency and fault tolerance.  
Kim et al. developed NodeFinder to measure Ethereum \gls{p2p} networks during the \gls{pow} era, finding that as of 2018, most nodes were in the U.S.\ (43.2\%), China (12.9\%), and Germany (5.2\%)~\cite{kim2018measuring}. 
Similar inference methods have been applied to the PoS era~\cite{bostoen2024estimating}, showing continued concentration in a few countries.  Online dashboards~\cite{ethernodes2025countries,ccaf2025ethereum} report real-time %
Ethereum node distribution %
and, as of December 2025, identify the United States and Germany as dominant. 
Heimbach et al.\ further developed a deanonymization technique that located over 15\% of validators, with more than 85\% situated in Europe and North America~\cite{heimbach2025deanonymizing}.
Kiffer et al.\ measured the \gls{p2p} infrastructure of 36 public blockchain networks, including Ethereum, and found that node deployment is highly geographically concentrated, with most nodes located in the United States and Western Europe~\cite{kiffer2025multiple}.

\parhead{\gls{abm} in blockchain}
\gls{abm} has been widely adopted to model the strategic behavior of blockchain participants. %
Kraner et al. %
modeled Ethereum \gls{pos} to analyze how network topology and latency influence consensus performance and stability~\cite{kraner2023agent}. {\"O}z et al. examined the impact of waiting games on consensus stability, showing that delay strategies can be profitable without undermining consensus when widely adopted~\cite {oz_time_2023}. %
Other works used \gls{abm} to study builder behaviors in MEV-Boost auctions~\cite{wu2024strategic,wu2024competition}, assess design proposals such as Execution Tickets~\cite{stichler2025galaxy}, and analyze how staking and inflation policies shape token concentration in DAO and DeFi governance~\cite{cong2024agent}.

\section{Conclusion}
In this paper, we develop a unified formal and computational framework to study how Ethereum's block-building paradigms interact with validator and information-source distributions to shape geographical positioning incentives. Through analytical characterization and controlled simulations, we show that Ethereum's block-building architecture is not geographically neutral. Both paradigms generate location-dependent payoffs and incentives to relocate closer to payoff-relevant parties in order to reduce propagation delays, although through different underlying mechanisms. Asymmetric access to information sources and suppliers further strengthens incentives toward geographical concentration.

We also demonstrate that consensus parameters, such as attestation thresholds and slot times, modulate latency sensitivity and can act as protocol-level levers that amplify or dampen these effects. Together, our results highlight that geographical decentralization is an emergent outcome of protocol design choices rather than a purely infrastructural concern. Finally, we discuss the implications of these findings for ongoing protocol evolution and outline potential mitigation directions aimed at preserving Ethereum's decentralization and security properties.

\begin{acks}
The authors thank the anonymous reviewers and the shepherd for their constructive comments and guidance.
The authors also thank Quintus Kilbourn, Lioba Heimbach, Thomas Thiery, Devan Mitchem, Akaki Mamageishvili, Istv\'an Andr\'as Seres, Bruno Mazorra, Christoph Schlegel, and Luis Correia for helpful discussions and comments. Fei Wu acknowledges support from a Flashbots research grant.
Sen Yang and Fan Zhang acknowledge support from a separate grant from Flashbots.
\end{acks}

\bibliographystyle{ACM-Reference-Format}
\bibliography{bibliography}

\appendix
\section{Omitted Proofs}
\label{sec:proofs}
\begin{proof}[Proof of \Cref{lem:tau-monotone}]
The optimal release time is $\tau_r^\star = \sup\{\tau : \zeta_r(\tau) \geq \gamma\}$. Since $\zeta_{r_1}(\tau) \geq \zeta_{r_2}(\tau)$ for all $\tau$, the set $\{\tau : \zeta_{r_1}(\tau) \geq \gamma\}$ contains $\{\tau : \zeta_{r_2}(\tau) \geq \gamma\}$, so its supremum is weakly larger.   
\end{proof}

\begin{proof}[Proof of \Cref{lem:payoff-monotone}]
Under the large-committee approximation, the proposer chooses the latest feasible release time, and the payoff is the effective value evaluated at that release time. Thus
\[
W_L(r)=\bar V_r(\tau_r^\star),\qquad W_E(I;r)=\bar V_I(\tau_I^\star(r);r).
\]

Since the effective value functions are non-decreasing in $\tau$, the payoff is non-decreasing in the corresponding optimal release time. Under the linear specification,
\[
\frac{d}{d\tau}\bar V_r(\tau)=\sum_{I\in\mathcal{I}_{\mathrm{signal}}}a_I>0,
\qquad
\frac{d}{d\tau}\bar V_I(\tau;r)=a_I>0,
\]
so the monotonicity is strict.

\end{proof}

\begin{proof}[Proof of \Cref{thm:latency-payoff}~(a)]
Under local block building, the proposer's payoff depends on delays to signal sources (for value accrual) and to attesters (for canonicalization). We show that both channels contribute. First, consider the value channel. If
\[
d(r_1,r(I))\preceq_{\mathrm{st}} d(r_2,r(I))
\qquad
\text{for every } I\in \mathcal{I}_{\mathrm{signal}},
\]
because each $V_I$ is non-decreasing,
\[
\mathbb{E}\!\left[V_I\!\left(\tau-d(r_1,r(I))\right)\right]
\geq
\mathbb{E}\!\left[V_I\!\left(\tau-d(r_2,r(I))\right)\right].
\]
Summing over signal sources gives
\[
\bar V_{r_1}(\tau) \geq \bar V_{r_2}(\tau)
\qquad
\text{for all } \tau.
\]

Next, consider the timing channel. For attester $a$,
\[
q_a(\tau;r)=F_{d(r,r(a))}(\tau_{\mathrm{cut}}-\tau).
\]
If
\[
d(r_1,r(a))\preceq_{\mathrm{st}} d(r_2,r(a)),
\]
then
\[
q_a(\tau;r_1)\ge q_a(\tau;r_2)
\qquad
\text{for all } \tau.
\]
Averaging over attesters yields
\[
\zeta_{r_1}(\tau)\ge \zeta_{r_2}(\tau)
\qquad
\text{for all } \tau.
\]
By \Cref{lem:tau-monotone},
\[
\tau_{r_1}^\star\ge \tau_{r_2}^\star.
\]
Therefore,
\[
W_L(r_1)=\bar V_{r_1}(\tau_{r_1}^\star)
\ge \bar V_{r_1}(\tau_{r_2}^\star)
\ge \bar V_{r_2}(\tau_{r_2}^\star)
= W_L(r_2),
\]
where the first inequality uses \Cref{lem:payoff-monotone} and the second uses the value comparison above.
\end{proof}

\begin{proof}[Proof of \Cref{thm:latency-payoff}~(b)]
Fix a supplier $I$ and a release time $\tau$. By assumption,
\[
d(r_1,r(I))\preceq_{\mathrm{st}} d(r_2,r(I)).
\]
Since $V_I$ is non-decreasing,
\[
\bar V_I(\tau;r_1)\ge \bar V_I(\tau;r_2).
\]

For dissemination, the timely-attestation probability for attester $a$ is
\[
q_a(\tau;I,r)
=
F_{d(r,r(I))+d(r(I),r(a))}(\tau_{\mathrm{cut}}-\tau).
\]
By the additive model and independence assumption, improving the proposer--supplier hop from $r_2$ to $r_1$ preserves first-order stochastic dominance of the end-to-end path to each attester, thus,
\[
q_a(\tau;I,r_1)\ge q_a(\tau;I,r_2)
\qquad
\text{for all } a,\tau.
\]
Averaging over attesters yields
\[
\zeta_I(\tau;r_1)\ge \zeta_I(\tau;r_2)
\qquad
\text{for all } \tau.
\]
By \Cref{lem:tau-monotone},
\[
\tau_I^\star(r_1)\ge \tau_I^\star(r_2).
\]
Therefore,
\[
W_E(I;r_1)=\bar V_I(\tau_I^\star(r_1);r_1)
\ge \bar V_I(\tau_I^\star(r_2);r_1)
\ge \bar V_I(\tau_I^\star(r_2);r_2)
= W_E(I;r_2),
\]
where the first inequality uses \Cref{lem:payoff-monotone} and the second uses the value comparison above.
Since this holds for every supplier $I$,
\[
W_E(r_1)=\max_I W_E(I;r_1)\ge \max_I W_E(I;r_2)=W_E(r_2).
\]
\end{proof}

\begin{proof}[Proof of \Cref{prop:signal-scaling}]
Under local block building, \Cref{rem:phi} gives $\phi=\tau$, so each proposer--source delay enters the value term only once. With equal attester delays,
\[
\tau_{r_1}^\star=\tau_{r_2}^\star=:\tau^\star.
\]
Hence
\[
W_L(r_1)-W_L(r_2)
=
\sum_{I\in\mathcal{I}_{\mathrm{signal}}}
a_I\Bigl(\mathbb{E}[d(r_2,r(I))]-\mathbb{E}[d(r_1,r(I))]\Bigr)
=
\delta\sum_{I\in\mathcal{I}_{\mathrm{signal}}}a_I.
\]
If all $a_I=a$, this simplifies to
\[
W_L(r_1)-W_L(r_2)=a\,\delta\,|\mathcal{I}_{\mathrm{signal}}|.
\]
\end{proof}

\begin{proof}[Proof of \Cref{prop:supplier-selection}]
Fix a supplier $I$. By \Cref{rem:phi}, under external block building
\[
V_I\!\left(\tau-d(r,r(I))\right)=V_I\!\left(\phi-2d(r,r(I))\right),
\]
so proposer--supplier latency enters the value term twice.

At a fixed commit time $\tau$, reducing the proposer--supplier delay by $\delta$ increases the effective value by exactly $a_I\delta$ under the linear specification. Equivalently, at a common attester-facing release time $\phi$, the same delay reduction would increase the value argument by $2\delta$, which explains the upper bound below.

In the optimization problem stated in terms of commit time $\tau$, the extra gain beyond $a_I\delta$ depends on how much later the proposer can commit while ensuring timely attestation. Since the end-to-end proposer--supplier--attester delay also shifts left by $\delta$,
\[
0 \le \tau_I^\star(r_1)-\tau_I^\star(r_2)\le \delta.
\]
Hence
\[
W_E(I;r_1)-W_E(I;r_2)
=
a_I\bigl(\tau_I^\star(r_1)-\tau_I^\star(r_2)\bigr)+a_I\delta,
\]
which implies
\[
a_I\delta \le W_E(I;r_1)-W_E(I;r_2)\le 2a_I\delta.
\]
If $\tau_I^\star(r_2)+\delta\le \tau_{\mathrm{cut}}$, then the feasible frontier shifts by exactly $\delta$, so equality holds at the upper bound. Finally,
\[
W_E(r_1)-W_E(r_2)
=
\max_I W_E(I;r_1)-\max_I W_E(I;r_2)
\le
\max_I\bigl(W_E(I;r_1)-W_E(I;r_2)\bigr),
\]
so
\[
W_E(r_1)-W_E(r_2)\le 2\delta \max_{I\in\mathcal{I}_{\mathrm{supplier}}} a_I.
\]
Therefore, the gain is controlled by the selected supplier and does not scale with the number of suppliers.
\end{proof}

\begin{proof}[Proof of \Cref{prop:opposing-placement}~(a)]
Let the $M$ signal sources be placed at regions $x_1,\dots,x_M$. Because source placement does not affect attester delays, the optimal release times $\tau_{r_s}^\star$ and $\tau_{r'}^\star$ are fixed across all source placements. With an identical source slope $a$, the local payoff gap is
\[
W_L(r_s)-W_L(r')
=
aM(\tau^\star_{r_s}-\tau^\star_{r'})
+
a\sum_{j=1}^M
\Bigl(\mathbb{E}[d(r',x_j)]-\mathbb{E}[d(r_s,x_j)]\Bigr).
\]
The first term is independent of source placement. By assumption,
\[
\mathbb{E}[d(r',r_s)]-\mathbb{E}[d(r_s,r_s)]
\ge
\mathbb{E}[d(r',x)]-\mathbb{E}[d(r_s,x)]
\qquad
\text{for every } x\in \mathcal{R},
\]
so each summand is maximized by choosing $x_j=r_s$. Therefore, the total payoff gap is maximized when all $M$ signal sources are concentrated in $r_s$. Since source placement does not affect attester delays and $r_s$ is at least as well connected to attesters as $r'$, we have
\[
\zeta_{r_s}(\tau)\ge \zeta_{r'}(\tau)
\qquad
\text{for all } \tau,
\]
so \Cref{lem:tau-monotone} gives $\tau_{r_s}^\star\ge \tau_{r'}^\star$. Thus, this concentrated placement amplifies migration incentives toward $r_s$ relative to dispersed placements.
\end{proof}

\begin{proof}[Proof of \Cref{prop:opposing-placement}~(b)]
Fix a supplier $I^\star$ located in $r_s$, and let $G(\delta)$ denote the payoff gain from a uniform reduction $\delta$ in the proposer--supplier delay while continuing to use supplier $I^\star$. 

At any fixed release time, reducing proposer--supplier delay by $\delta$ increases the effective value by $a_{I^\star}\delta$. Moreover, a larger shift parameter weakly enlarges the feasible set, so the corresponding optimal release time is non-decreasing in $\delta$ by \Cref{lem:tau-monotone}. Hence
\[
G(\delta)
=
a_{I^\star}\delta
+
a_{I^\star}\bigl(\tau_{I^\star}^\star(\delta)-\tau_{I^\star}^\star(0)\bigr)
\]
is non-decreasing in $\delta$. Therefore, a supplier located in a more remote region induces a stronger migration incentive toward that region.
\end{proof}

\begin{proof}[Proof of \Cref{prop:slot-duration}]
Under the stated assumptions, reducing $\tau_{\mathrm{cut}}$ shifts every region's optimal release time left by the same amount and leaves all latency terms unchanged. Under local block building, payoff takes the form
\[
W_L(r)=a|\mathcal{I}_{\mathrm{signal}}|\tau_{\mathrm{cut}}-C_r
\]
for a region-specific constant $C_r$ independent of $\tau_{\mathrm{cut}}$. Under external block building, payoff takes the form
\[
W_E(r)=a\tau_{\mathrm{cut}}-C_r'
\]
for a region-specific constant $C_r'$ independent of $\tau_{\mathrm{cut}}$, provided the maximizing supplier does not change. Thus shortening the slot subtracts the same constant from every region's payoff in either paradigm. Pairwise payoff differences are unchanged, while the mean payoff decreases. 
\end{proof}

\section{List of Google Cloud Platform Regions}
\label{sec:gcp-regions}

\begin{figure}[h]
    \centering
    \includegraphics[width=0.9\linewidth]{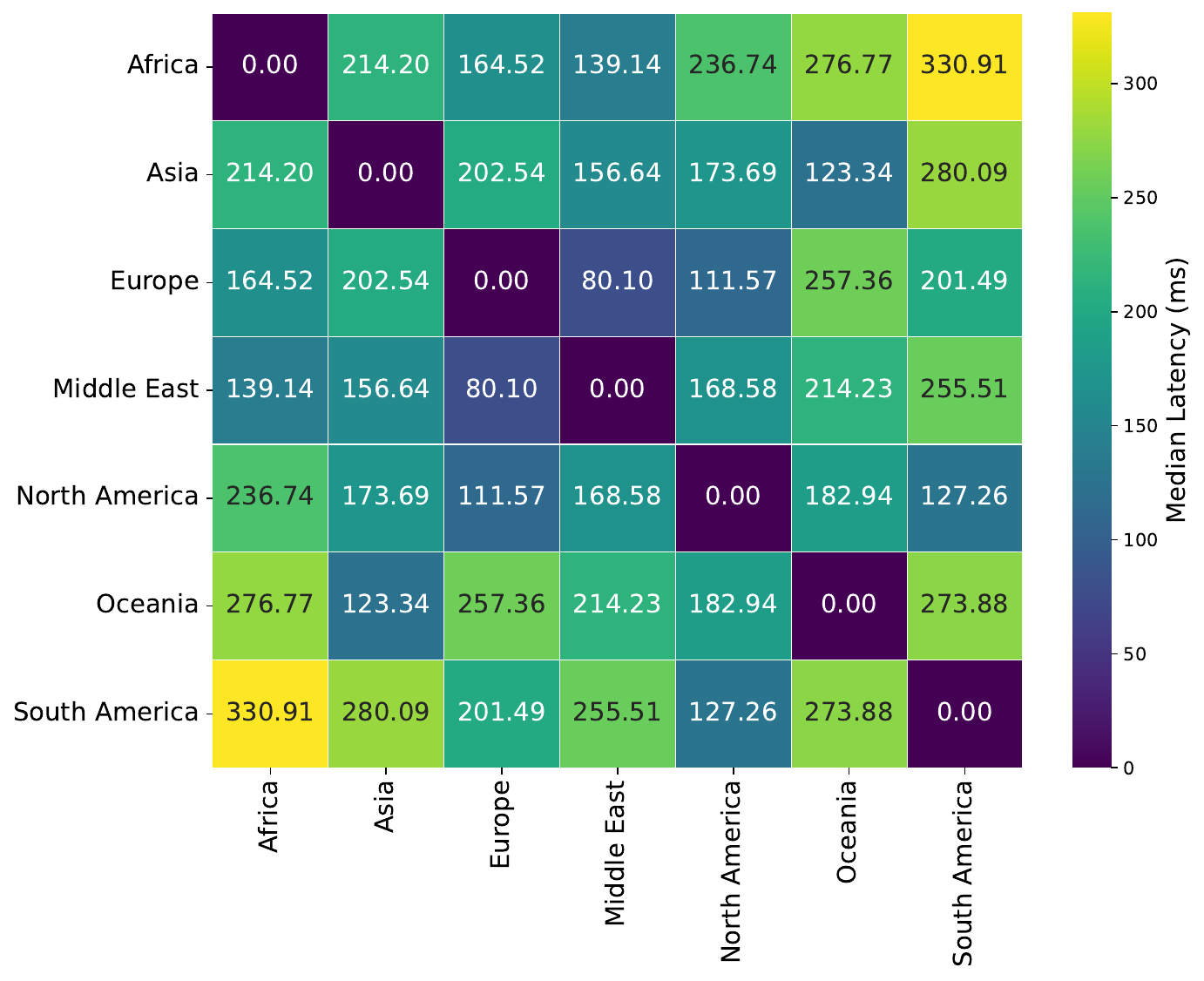}
    \caption{Heatmap of median latency between macro-regions.}
    \Description{Heatmap of median latency between macro-regions.}
    \label{fig:latency-heatmap}
\end{figure}

In this section, we present the list of Google Cloud Platform (GCP) regions used in our paper, along with their physical locations, as shown in~\Cref{tab:gcp-regions}.
We exclude two GCP regions---me-central2 (Dammam, Saudi Arabia) and northamerica-south1 (Queretaro, Mexico)---because latency data between these regions and the others was not available~\cite{google2025lookerstudio}.
\Cref{fig:latency-heatmap} presents the heatmap of median latency between each pair of macro-regions.
For example, the median latency between Asia and Europe is computed as the median value of all latencies between GCP regions in Asia and GCP regions in Europe.
Among the seven macro-regions, North America has the best average median latency to other regions (142.97 ms), while South America has the worst (209.88 ms).

{\small
\begin{longtable}{lll}
\caption{GCP regions and their physical locations.}
\label{tab:gcp-regions}\\
\toprule
\textbf{Macro-Region} & \textbf{GCP Region} & \textbf{Location} \\
\midrule
\endfirsthead

\toprule
\textbf{Macro-Region} & \textbf{GCP Region} & \textbf{Location} \\
\midrule
\endhead

\bottomrule
\endfoot

Africa  & africa-south1 & Johannesburg, South Africa \\
\midrule
\multirow{9}{*}{Asia} & asia-east1 & Changhua County, Taiwan \\
~ & asia-east2 & Hong Kong, China \\
~ & asia-northeast1 & Tokyo, Japan \\
~ & asia-northeast2 & Osaka, Japan \\
~ & asia-northeast3 & Seoul, South Korea \\
~ & asia-south1 & Mumbai, India \\
~ & asia-south2 & Delhi, India \\
~ & asia-southeast1 & Singapore \\
~ & asia-southeast2 & Jakarta, Indonesia \\
\midrule
\multirow{2}{*}{Oceania} & australia-southeast1 & Sydney, Australia \\
~ & australia-southeast2 & Melbourne, Australia \\
\midrule
\multirow{13}{*}{Europe} & europe-central2 & Warsaw, Poland \\
~ & europe-north1 & Hamina, Finland \\
~ & europe-north2 & Stockholm, Sweden \\
~ & europe-southwest1 & Madrid, Spain \\
~ & europe-west1 & St.\ Ghislain, Belgium \\
~ & europe-west2 & London, UK \\
~ & europe-west3 & Frankfurt, Germany \\
~ & europe-west4 & Eemshaven, Netherlands \\
~ & europe-west6 & Zurich, Switzerland \\
~ & europe-west8 & Milan, Italy \\
~ & europe-west9 & Paris, France \\
~ & europe-west10 & Berlin, Germany \\
~ & europe-west12 & Turin, Italy \\
\midrule
\multirow{2}{*}{Middle East} & me-central1 & Doha, Qatar \\
~ & me-west1 & Tel Aviv, Israel \\
\midrule
\multirow{2}{*}{South America} & southamerica-east1 & Sao Paulo, Brazil \\
~ & southamerica-west1 & Santiago, Chile \\
\midrule
\multirow{11}{*}{North America} & northamerica-northeast1 & Montreal, Canada \\
~ & northamerica-northeast2 & Toronto, Canada \\
~ & us-central1 & Council Bluffs, Iowa, USA \\
~ & us-east1 & Moncks Corner, SC, USA \\
~ & us-east4 & Ashburn, Virginia, USA \\
~ & us-east5 & Columbus, Ohio, USA \\
~ & us-south1 & Dallas, Texas, USA \\
~ & us-west1 & The Dalles, Oregon, USA \\
~ & us-west2 & Los Angeles, California, USA \\
~ & us-west3 & Salt Lake City, Utah, USA \\
~ & us-west4 & Las Vegas, Nevada, USA \\
\end{longtable}
}

\section{List of Symbols}

\begin{table}[H]
\centering
\caption{List of symbols.}
\begin{tabularx}{\textwidth}{@{}l X@{}}
\toprule
\textbf{Symbol} & \textbf{Description} \\
\midrule

\textbf{Geography and Latency}\\
\arrayrulecolor{gray}
\cmidrule{1-1}
\arrayrulecolor{black}
$\mathcal{R} = \{r_1,\dots,r_m\}$ & Set of geographical regions \\
$d(r_j,r_k)$ & Message propagation latency from region $r_j$ to $r_k$, drawn from a log-normal distribution \\
$\mu_{jk}$ & Mean of $\ln d(r_j,r_k)$ \\

\midrule
\textbf{Validators and Consensus}\\
\arrayrulecolor{gray}
\cmidrule{1-1}
\arrayrulecolor{black}
$\mathcal{V} = \{v_1,\dots,v_P\}$ & Set of validators (agents) \\
$n$ & Slot index, $n \in \{1,\dots,N\}$ \\
$p_n$ & Proposer selected for slot $n$ \\
$r(p_n)$ & Region of the proposer of slot $n$ \\
$a \in \mathcal{A}_n$ & An attestor in the set of attestors assigned to slot $n$ \\
$r(a)$ & Region of attester $a$ \\
$\gamma$ & Fraction of attestation required for canonicalization \\
$\Delta$ & Slot duration \\
$\tau$ & Block release time within a slot \\
$\tau_{\text{cut}}$ & Cutoff time for timely attestations \\
$c$ & Cost of migration across regions \\

\midrule
\textbf{Information Sources}\\
\arrayrulecolor{gray}
\cmidrule{1-1}
\arrayrulecolor{black}
$\mathcal{I}$ & Set of information sources \\
$\mathcal{I}_{\text{signal}}$ & Set of signal sources (used in the local block building paradigm) \\
$\mathcal{I}_{\text{supplier}}$ & Set of block suppliers (used in the external block building paradigm) \\
$r(I)$ & Region of signal source or block supplier $I$ \\
$V_I(t) = a_I t + b_I$ & Value available from source or supplier $I$ at time $t$ \\
$V_r(\tau)$ & Aggregate block value for a proposer located in region $r$ at release time $\tau$ \\

\midrule
\textbf{Attestation and Payoffs}\\
\arrayrulecolor{gray}
\cmidrule{1-1}
\arrayrulecolor{black}
$q_a(\tau;r)$ & Probability that attester $a$ votes on time when a block is released at $(r,\tau)$ \\
$S(\tau;r)$ & Number of timely attestations for a block released at $(r,\tau)$ \\
$\Pi_r(\tau)$ & Probability that a block released at $(r,\tau)$ becomes canonical \\
$\tau_r^\star$ & Optimal block release time for a proposer in region $r$ \\
$W(r)$ & Expected payoff for a proposer located in region $r$ \\
$\tau_I^\star$ & Optimal release time when using block supplier $I$ \\
$W(I)$ & Expected payoff when using block supplier $I$ \\
$\mathcal{W}$ & Set of best expected proposer payoffs across regions in a slot \\

\arrayrulecolor{black}
\bottomrule
\end{tabularx}
\label{tab:symbols}
\end{table}

\section{Experiments with Alternative $\sigma$ Values in the Log-Normal Latency Model}
\label{sec:different-sigmas}

This section performs sensitivity checks around the baseline configuration with $c=0$ by varying only the $\sigma$ parameter in the log-normal latency model. For each value of $\sigma$, we conduct 20 independent simulation runs with different random seeds and compare the mean trajectories across configurations. To maintain comparability, we use the same set of seeds for every $\sigma$ value. Specifically, we vary $\sigma$ over $\{0.2, 0.3, 0.4, 0.5, 0.6, 0.7, 0.8\}$, covering a broad range of latency variability.

\begin{figure}[thbp]
    \centering
    \includegraphics[width=0.8\linewidth]{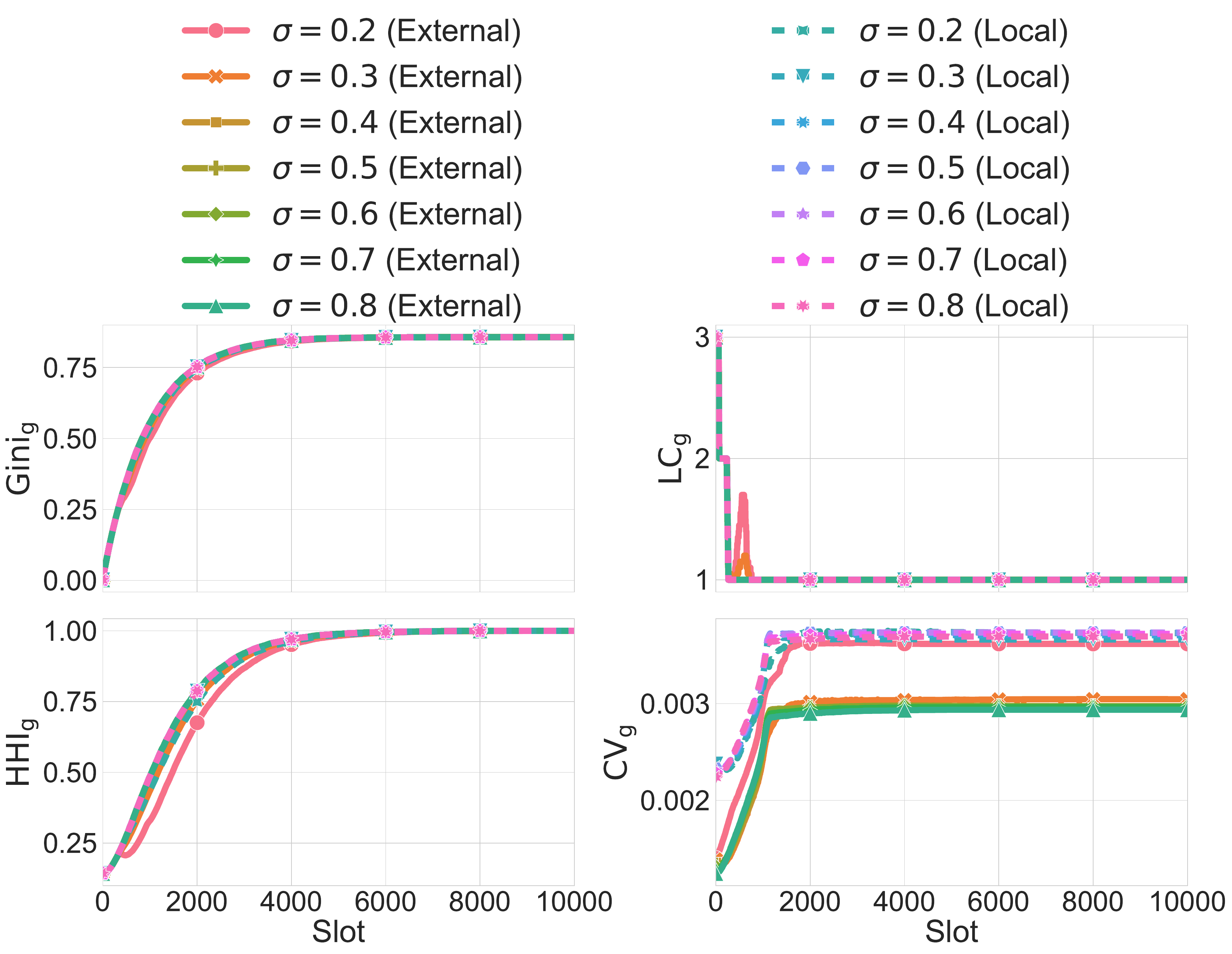}
    \caption{\emph{Setup: Homogeneous validators and information sources with $c = 0$.} Evolution of centralization metrics under Local and External block-building paradigms with varying $\sigma = \{0.2, 0.3, 0.4, 0.5, 0.6, 0.7, 0.8\}$.}
    \Description{}
    \label{fig:comparsion-cross-sigmas}
\end{figure}

\Cref{fig:comparsion-cross-sigmas} examines how latency variability affects centralization dynamics under both external and local block-building paradigms. Across all tested values, the trajectories of $\mathrm{Gini}_{\mathrm{g}}$, $\mathrm{HHI}_{\mathrm{g}}$, and $\mathrm{LC}_{\mathrm{g}}$ remain highly similar under both paradigms, with nearly identical convergence patterns and final levels. This suggests that the qualitative conclusions of the baseline configuration are robust to moderate changes in latency variability.
The main visible sensitivity appears in $\mathrm{CV}_{\mathrm{g}}$, where different $\sigma$ values lead to small but systematic shifts in the steady-state level. This is because, in our model, $\sigma$ determines how sharply the canonicalization probability drops near the release deadline. Smaller $\sigma$ makes this transition steeper, so small regional latency differences translate into larger differences in feasible release times, which in turn amplify payoff dispersion across regions.

Overall, these results show that our conclusions are robust to this modeling assumption over a broad range of $\sigma$ values, with only modest quantitative differences in $\mathrm{CV}_{\mathrm{g}}$.

\section{Experiments with Different Scales}
\label{sec:different-scales}

\begin{figure}[thbp]
    \centering
    \includegraphics[width=0.8\linewidth]{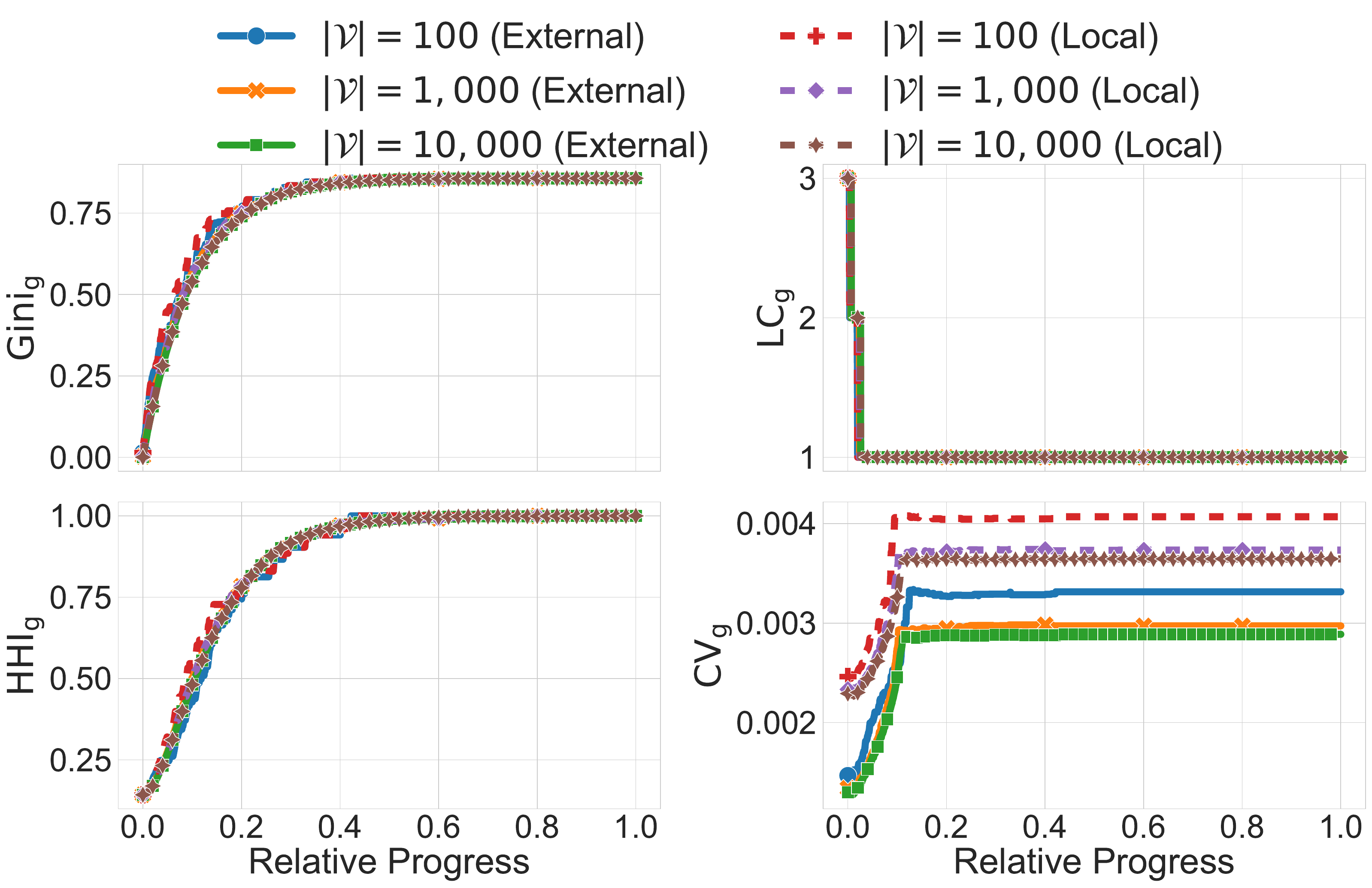}
    \caption{\emph{Setup: Homogeneous validators and information sources with $c = 0$.} Evolution of centralization metrics under Local and External block-building paradigms across different system scales, with $(|\mathcal{V}|, N) \in {(100, 1{,}000), (1{,}000, 10{,}000), (10{,}000, 100{,}000)}$. The x-axis shows normalized progress, defined as the current slot index divided by the total number of slots.}
    \Description{}
    \label{fig:comparison-cross-scale}
\end{figure}

We evaluated whether our results are sensitive to the scale of the simulation by running the same experiment (with homogeneous validators and information sources, and $c = 0$) under three configurations: $|\mathcal{V}| = 100$ validators with $N = 1{,}000$ slots; $|\mathcal{V}| = 1{,}000$ validators with $N = 10{,}000$ slots; and $|\mathcal{V}| = 10{,}000$ validators with $N = 100{,}000$ slots.
Because large-scale simulations are computationally expensive, \Cref{fig:comparison-cross-scale} presents a single run for each configuration; the purpose is to compare qualitative trajectory alignment across scales rather than estimate uncertainty.

While the absolute values of the per-slot metrics differ across scales, this is largely due to the different slot horizons. As shown in~\Cref{fig:comparison-cross-scale}, when we normalize time by the normalized relative progress within an experiment (i.e., the current slot index divided by the total number of slots), the trajectories align closely and exhibit the same qualitative trends across all three configurations.

We observe that the $\mathrm{CV}_{\mathrm{g}}$ exhibits slightly larger differences across scales compared to the other centralization metrics. This is mainly driven by our latency model, in which the probability that a block becomes canonical depends on a Poisson–binomial tail probability over timely attestations. When the attester (validator) population is small (e.g., $|\mathcal{V}|=100$), this Poisson–binomial aggregation exhibits stronger finite-size effects, so small differences in per-attester latency probabilities $q_a(\tau;r)$ translate into more visible differences in $\Pi_r(\tau)$ and, consequently, the resulting per-slot outcomes.
As the system scales to $|\mathcal{V}| = 1{,}000$ and $10{,}000$, this aggregation concentrates and the effect diminishes. Importantly, despite these scale-dependent differences in absolute $\mathrm{CV}_{\mathrm{g}}$ values, the relative ordering, temporal evolution, and overall qualitative trends remain consistent across all configurations, indicating that our conclusions are driven by the underlying dynamics rather than a particular system scale.

Simulations with larger validator populations are substantially more expensive. For $|\mathcal{V}| = 10{,}000$, the runtime of a single-slot simulation already exceeds one second, making a full simulation with $N = 100{,}000$ slots take more than a day to complete. We therefore adopt $|\mathcal{V}| = 1{,}000$ for most experiments in this paper, as it provides a practical balance between computational feasibility and statistical stability while preserving the same trends observed at larger scales.

\section{Marginal Benefit Distribution}
\label{sec:profit-gap}

\begin{figure}[htbp]
    \centering
    \includegraphics[width=0.7\linewidth]{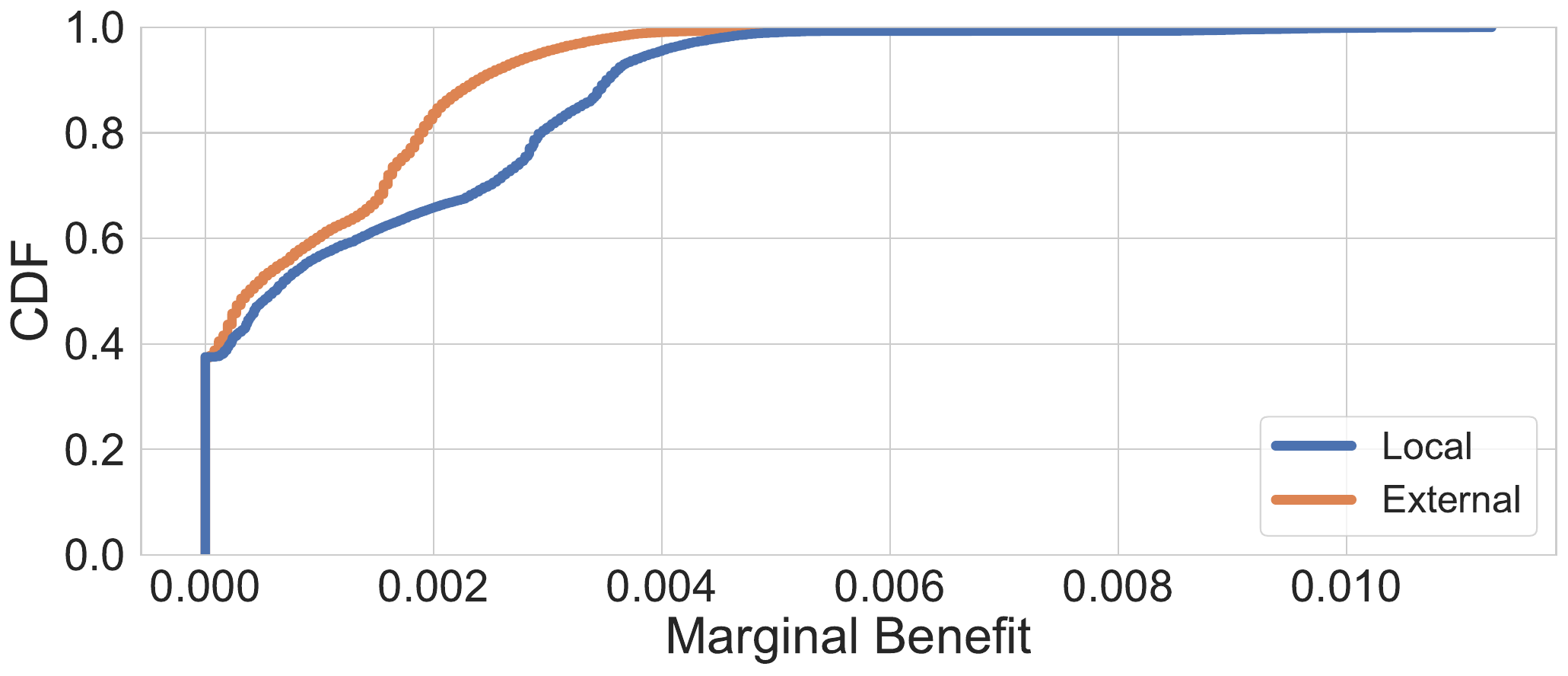}
    \caption{CDF of marginal benefit in the first \SI{1000}{} slots under the baseline configuration.}
    \Description{}
    \label{fig:profit-gap-cdf}
\end{figure}

\Cref{fig:profit-gap-cdf} shows the \gls{cdf} of marginal benefits when proposers move to the best region under the baseline configuration with migration cost $c=0$. 
The \gls{cdf} is pooled over the first \SI{1000}{} slots from each of 20 independent runs, where both validators and information sources are uniformly distributed across macro-regions and evenly within each macro-region's GCP regions.
Under the external block-building paradigm, about 80\% of migration marginal benefits are less than 0.002, which suggests that such migration could be prevented if the migration cost is 0.002.

\section{Migration Costs}
\label{sec:different-costs}
\begin{figure}[thbp]
    \centering
    \includegraphics[width=0.9\linewidth]{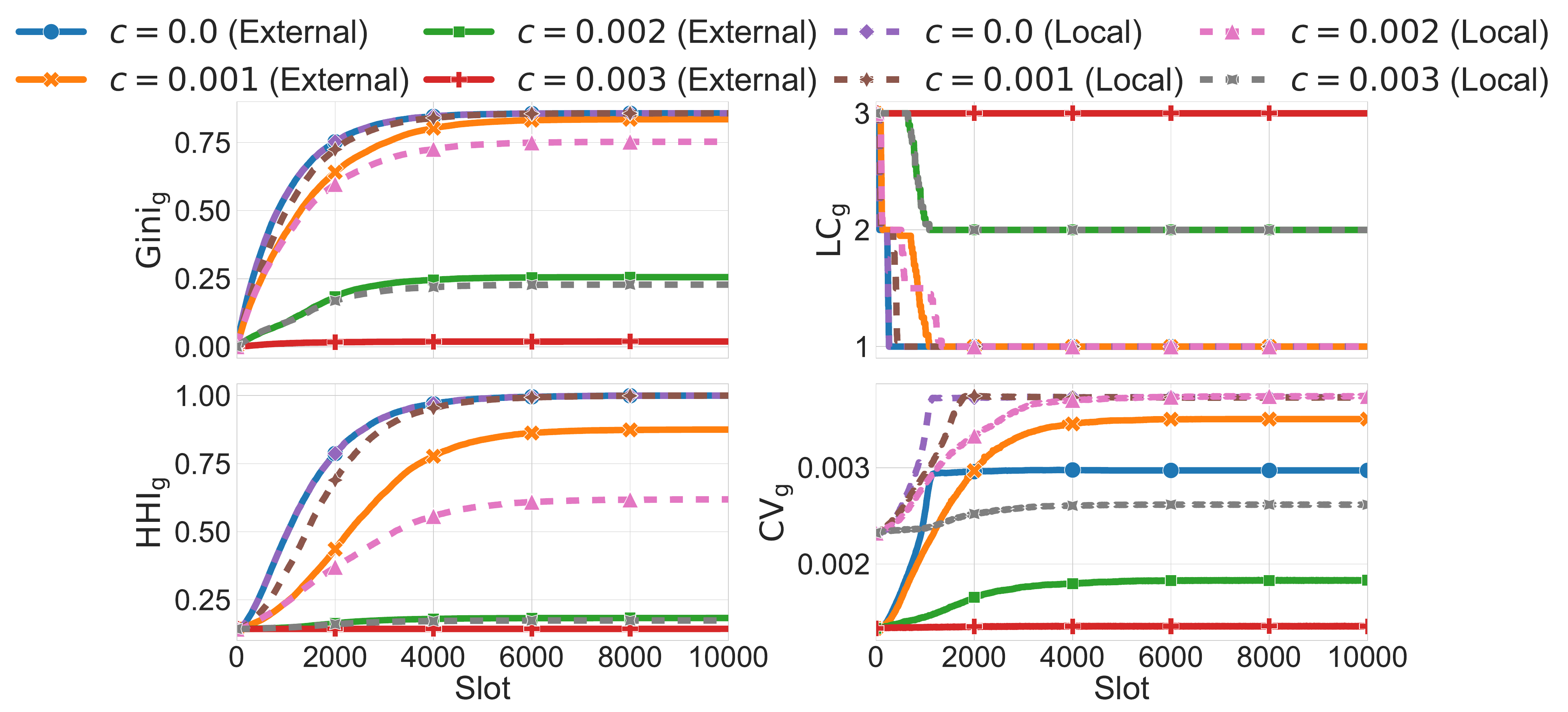}
    \caption{\emph{Setup: Homogeneous validators and information sources.} Evolution of centralization metrics under Local and External block-building paradigms with varying migration costs $c \in \{0, 0.001, 0.002, 0.003\}$.} 
    \Description{Setup: Homogeneous validators and information sources. Evolution of centralization metrics under Local and External block-building paradigms with varying migration costs $c \in \{0, 0.001, 0.002, 0.003\}$. }
    \label{fig:cost-metrics}
\end{figure}

We evaluate how migration costs affect both the degree and speed of geographical centralization. Using the baseline configuration in \Cref{sec:baseline} (homogeneous validators and information sources), we vary the cost from $0$ to $0.003$; at the upper end of this range, the cost removes over $70\%$ of the typical marginal relocation gain in both the local and external block-building paradigms (see \Cref{fig:profit-gap-cdf}). As shown in \Cref{fig:cost-metrics}, higher costs predictably dampen validator mobility and mitigate centralization. This migration-suppressing effect is consistently stronger under the external block-building paradigm than under the local block-building paradigm, reflecting \Cref{sec:baseline}: under the local block-building paradigm, location directly shifts marginal value via multi-source aggregation, sustaining stronger incentives to co-locate even when migration is costly.

\section{Baseline Configuration: Validator Convergence Locus}\label{sec:convergence_locus}

\begin{figure}[hbtp]
\centering
\includegraphics[width=0.99\linewidth]{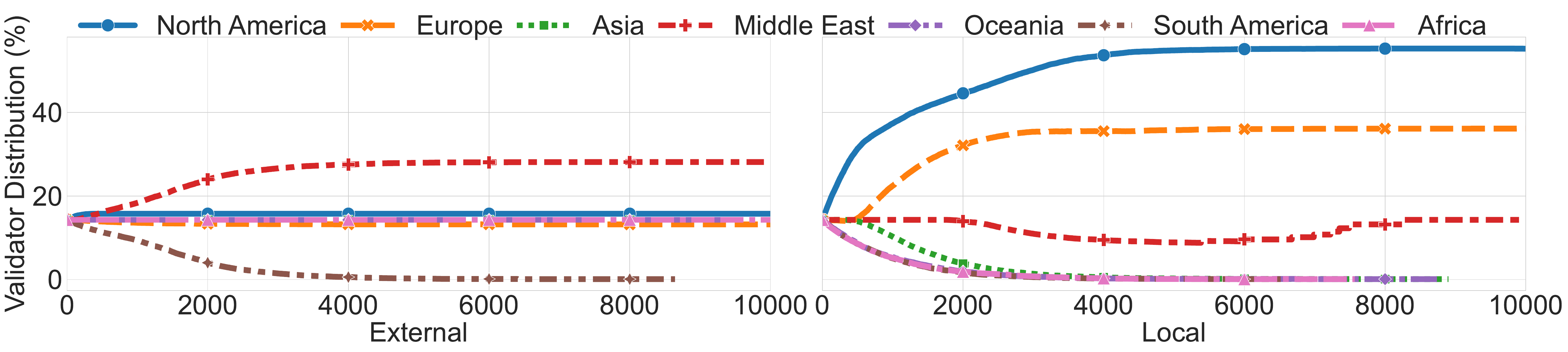}
\caption{\emph{Baseline configuration (homogeneous validators and information sources).} Evolution of validator distribution aggregated by macro-regions under Local and External block-building paradigms.
Early termination of a line indicates that no validators remain located within the corresponding macro-region.}
\Description{}
\label{fig:region-distribution-baseline}
\end{figure}

\Cref{fig:region-distribution-baseline} illustrates how validators redistribute across macro-regions over time in the baseline simulation of \Cref{sec:baseline}, starting from an initially uniform geographical distribution.
Each curve shows the mean validator share of a macro-region across 20 independent simulation runs. 
This visualization highlights not only the overall degree of centralization but also \emph{where} validators ultimately converge, providing insight into which regions offer systematic advantages.

Under the external block-building paradigm, migration yields only modest gains, as discussed in~\Cref{sec:baseline}.
A small fraction of validators initially relocates to North America, slightly altering the validator distribution.
This shift not only benefits validators in North America but also alters the attester distribution, indirectly improving supplier--attester latency for other macro-regions.
Owing to its geographic position, the Middle East exhibits relatively low average latency to multiple macro-regions (see~\Cref{fig:latency-heatmap}) and thus gains a slight relative advantage as the distribution evolves.
As a result, validators initially located in South America tend to relocate to the Middle East, while most other regions experience weak migration incentives and largely retain their initial shares.

Under the local block-building paradigm, migration incentives are stronger because validator location directly affects both value accumulation and propagation to attesters. Validators initially concentrate most strongly in North America, while Europe also attracts validators early in the migration process due to favorable latency conditions. As migration proceeds, validators in higher-latency regions gradually leave their initial locations, and most of this movement is absorbed by North America and Europe. North America ultimately becomes the dominant hub, while Europe stabilizes as a secondary concentration center. Migration costs, however, prevent full convergence, leaving a non-trivial fraction of validators remaining in the Middle East.

\section{Joint Heterogeneity: Validator Convergence Locus}
\label{sec:full-hetero-cluster}
\begin{figure}[thbp]
    \centering
    \includegraphics[width=0.99\linewidth]{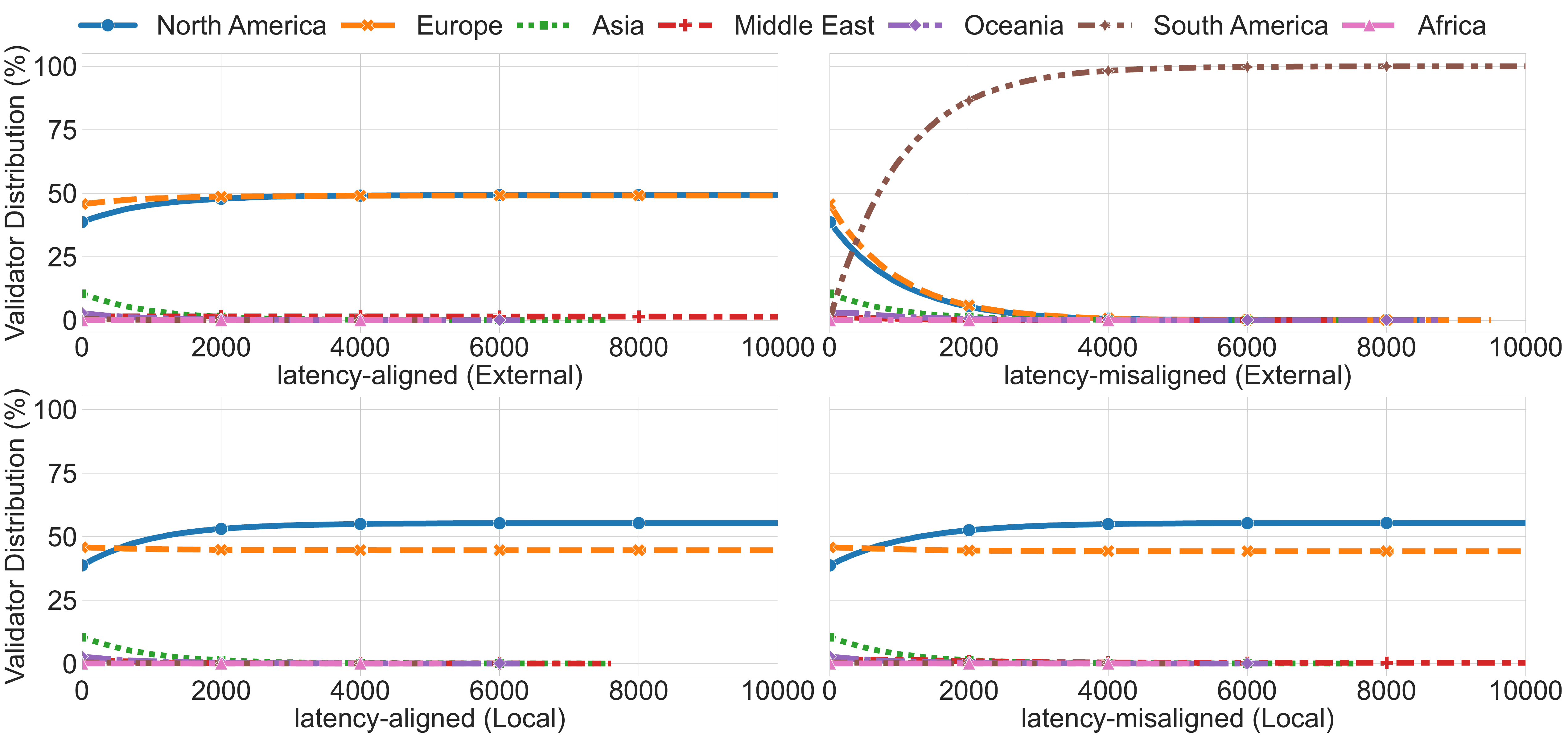}
    \caption{\emph{Joint heterogeneity (validators and information sources).} Evolution of validator distribution aggregated by macro-regions under Local and External block-building paradigms for latency-aligned and latency-misaligned information-source placements with a heterogeneous validator distribution. Early termination of a line indicates that no validators remain located within the corresponding macro-region.}
    \Description{}
    \label{fig:region-distribution-joint}
\end{figure}

\Cref{fig:region-distribution-joint} illustrates how validators redistribute across macro-regions over time in the joint-heterogeneity setting of \Cref{sec:se3}, combining latency-aligned and latency-misaligned information-source placements with the empirically observed validator distribution, which is initially concentrated in Europe and North America.

Consistent with the results in \Cref{sec:se4}, starting from an already concentrated validator distribution results in rapid stabilization of validator locations, with little subsequent redistribution across regions. Consequently, only a small fraction of validators relocate under most configurations: modest migration toward North America occurs under both latency-aligned and latency-misaligned placements in the local block-building paradigm, as well as under the latency-aligned configuration in the external block-building paradigm. The existing concentration of validators in North America and Europe creates favorable propagation conditions to attesters, which continue to outweigh incentives to migrate elsewhere, even when information sources are located in other regions, under the local block-building paradigm.

A distinct pattern emerges under the external block-building paradigm when information sources are located in high-latency regions, as shown in the upper-right panel of \Cref{fig:region-distribution-joint}. In this configuration, validator mass shifts away from incumbent hubs and converges toward the supplier region, despite the initial concentration in North America and Europe. This occurs because block dissemination originates from the supplier, so proximity to other validators provides little benefit, while proximity to the supplier becomes the dominant factor shaping validator placement. As a result, validators ultimately concentrate in a single supplier region, leading to renewed geographical centralization.

\section{Uncertainty Estimates}
\label{sec:uncertainty-estimates}

For each experimental configuration, we perform 20 independent simulation runs with different random seeds. Each reported statistic is computed at the run level from the final-slot outcome and then averaged across runs. We report 95\% CI obtained from the empirical distribution of the 20 run-level outcomes using percentile bootstrap resampling. These uncertainty estimates show that the reported simulation outcomes are not driven by a single run, but are stable across independent realizations of the stochastic simulation process.

In each table, every cell reports the mean together with its 95\% CI, formatted as \texttt{mean [lower bound, upper bound]}. Degenerate intervals indicate that all runs converged to the same final value up to the reported precision. \Cref{tab:final-metrics-baseline} reports the final-slot mean values and 95\% CI for the baseline configuration.
\Cref{tab:final-metrics-hetero-info,tab:final-metrics-hetero-validators,tab:final-metrics-hetero-both} report the corresponding statistics for the heterogeneous-information, heterogeneous-validator, and jointly heterogeneous settings in \Cref{sec:se1,sec:se2,sec:se3}, respectively.
\Cref{tab:final-metrics-different-gammas,tab:final-metrics-eip7782} report the corresponding statistics for the alternative consensus-parameter settings in \Cref{sec:se4}.
\Cref{tab:final-metrics-sigmas,tab:final-metrics-cost} report the corresponding statistics for different $\sigma$ values and migration-cost settings in \Cref{sec:different-sigmas,sec:different-costs}, respectively.

Across all tables, the zero-width 95\% CIs for $\mathrm{LC}_{\mathrm{g}}$ arise because $\mathrm{LC}_{\mathrm{g}}$ is a discrete threshold-based metric. Although stochastic realizations differ across runs, the final validator distribution under a fixed configuration typically remains in the same threshold regime, so the minimum number of macro regions required to cover one-third of validators is identical across runs. As a result, all runs yield the same final $\mathrm{LC}_{\mathrm{g}}$ value, although this value differs across configurations.

The near-zero width 95\% CIs for $\mathrm{CV}_{\mathrm{g}}$ reflect the stability of the final cross-region payoff profile across runs. Since $\mathrm{CV}_{\mathrm{g}}$ is defined as the coefficient of variation of the best proposer payoffs across regions within a slot, runs under the same configuration typically converge to similar final validator distributions. This in turn yields nearly identical regional payoff vectors and thus nearly identical final $\mathrm{CV}_{\mathrm{g}}$ values across runs, with some intervals appearing as zero-width after rounding.

\done%

\begin{table}[ht]
\centering
\caption{Final-slot metrics under baseline configurations.}
\label{tab:final-metrics-baseline}
\resizebox{\textwidth}{!}{
\begin{tabular}{lcccc}
\toprule
\textbf{Configuration} & $\boldsymbol{\mathrm{Gini}_{\mathrm{g}}}$& $\boldsymbol{\mathrm{LC}_{\mathrm{g}}}$ & $\boldsymbol{\mathrm{HHI}_{\mathrm{g}}}$ & $\boldsymbol{\mathrm{CV}_{\mathrm{g}}}$ \\
\midrule
External & 0.2558 [0.2551, 0.2566] & 2.0000 [2.0000, 2.0000] & 0.1829 [0.1822, 0.1835] & 0.0018 [0.0018, 0.0018] \\
Local & 0.7527 [0.7196, 0.7859] & 1.0000 [1.0000, 1.0000] & 0.6187 [0.5141, 0.7233] & 0.0037 [0.0037, 0.0038] \\
\bottomrule
\end{tabular}
}
\end{table}

\begin{table}[ht]
\centering
\caption{Final-slot metrics under the latency-aligned (LA) and latency-misaligned (LM) configurations with homogeneous validator distribution. \done}
\label{tab:final-metrics-hetero-info}
\resizebox{\textwidth}{!}{
\begin{tabular}{lcccc}
\toprule
\textbf{Configuration} & $\boldsymbol{\mathrm{Gini}_{\mathrm{g}}}$& $\boldsymbol{\mathrm{LC}_{\mathrm{g}}}$ & $\boldsymbol{\mathrm{HHI}_{\mathrm{g}}}$ & $\boldsymbol{\mathrm{CV}_{\mathrm{g}}}$ \\
\midrule
LA (External) & 0.8226 [0.8226, 0.8226] & 1.0000 [1.0000, 1.0000] & 0.7873 [0.7873, 0.7873] & 0.0026 [0.0026, 0.0026] \\
LM (External) & 0.8526 [0.8526, 0.8526] & 1.0000 [1.0000, 1.0000] & 0.9685 [0.9685, 0.9685] & 0.0039 [0.0039, 0.0039] \\
LA (Local) & 0.8571 [0.8571, 0.8571] & 1.0000 [1.0000, 1.0000] & 1.0000 [1.0000, 1.0000] & 0.0047 [0.0047, 0.0047] \\
LM (Local) & 0.8571 [0.8571, 0.8572] & 1.0000 [1.0000, 1.0000] & 0.9998 [0.9995, 1.0000] & 0.0033 [0.0033, 0.0033] \\
\bottomrule
\end{tabular}
}
\end{table}

\begin{table}[ht]
\centering
\caption{Final-slot metrics with heterogeneous validator distribution and homogeneous information sources.}
\label{tab:final-metrics-hetero-validators}
\resizebox{\textwidth}{!}{
\begin{tabular}{lcccc}
\toprule
\textbf{Configuration} & $\boldsymbol{\mathrm{Gini}_{\mathrm{g}}}$& $\boldsymbol{\mathrm{LC}_{\mathrm{g}}}$ & $\boldsymbol{\mathrm{HHI}_{\mathrm{g}}}$ & $\boldsymbol{\mathrm{CV}_{\mathrm{g}}}$ \\
\midrule
External & 0.7376 [0.7357, 0.7396] & 1.0000 [1.0000, 1.0000] & 0.5072 [0.5041, 0.5103] & 0.0028 [0.0028, 0.0028] \\
Local & 0.7348 [0.7311, 0.7385] & 1.0000 [1.0000, 1.0000] & 0.5045 [0.4998, 0.5093] & 0.0036 [0.0036, 0.0036] \\
\bottomrule
\end{tabular}
}
\end{table}

\begin{table}[ht]
\centering
\caption{Final-slot metrics under the latency-aligned (LA) and latency-misaligned (LM) configurations with heterogeneous validator distribution.}
\label{tab:final-metrics-hetero-both}
\resizebox{\textwidth}{!}{
\begin{tabular}{lcccc}
\toprule
\textbf{Configuration} & $\boldsymbol{\mathrm{Gini}_{\mathrm{g}}}$& $\boldsymbol{\mathrm{LC}_{\mathrm{g}}}$ & $\boldsymbol{\mathrm{HHI}_{\mathrm{g}}}$ & $\boldsymbol{\mathrm{CV}_{\mathrm{g}}}$ \\
\midrule
LA (External) & 0.7280 [0.7239, 0.7321] & 1.0000 [1.0000, 1.0000] & 0.4976 [0.4929, 0.5024] & 0.0028 [0.0028, 0.0028] \\
LM (External) & 0.8571 [0.8571, 0.8571] & 1.0000 [1.0000, 1.0000] & 1.0000 [1.0000, 1.0000] & 0.0045 [0.0045, 0.0045] \\
LA (Local) & 0.7295 [0.7278, 0.7312] & 1.0000 [1.0000, 1.0000] & 0.5060 [0.5047, 0.5074] & 0.0043 [0.0043, 0.0043] \\
LM (Local) & 0.7286 [0.7249, 0.7323] & 1.0000 [1.0000, 1.0000] & 0.5042 [0.4987, 0.5096] & 0.0028 [0.0028, 0.0029] \\
\bottomrule
\end{tabular}
}
\end{table}

\begin{table}[ht]
\centering
\caption{Final-slot metrics under different attestation-threshold settings.}
\label{tab:final-metrics-different-gammas}
\resizebox{\textwidth}{!}{
\begin{tabular}{lcccc}
\toprule
\textbf{Configuration} & $\boldsymbol{\mathrm{Gini}_{\mathrm{g}}}$& $\boldsymbol{\mathrm{LC}_{\mathrm{g}}}$ & $\boldsymbol{\mathrm{HHI}_{\mathrm{g}}}$ & $\boldsymbol{\mathrm{CV}_{\mathrm{g}}}$ \\
\midrule
$\gamma=1/2$ (External) & 0.0009 [0.0009, 0.0009] & 3.0000 [3.0000, 3.0000] & 0.1429 [0.1429, 0.1429] & 0.0011 [0.0011, 0.0011] \\
$\gamma=1/3$ (External) & 0.0009 [0.0009, 0.0009] & 3.0000 [3.0000, 3.0000] & 0.1429 [0.1429, 0.1429] & 0.0008 [0.0008, 0.0008] \\
$\gamma=2/3$ (External) & 0.2558 [0.2551, 0.2566] & 2.0000 [2.0000, 2.0000] & 0.1829 [0.1822, 0.1835] & 0.0018 [0.0018, 0.0018] \\
$\gamma=4/5$ (External) & 0.2486 [0.2482, 0.2490] & 2.0000 [2.0000, 2.0000] & 0.1778 [0.1776, 0.1780] & 0.0021 [0.0021, 0.0021] \\
$\gamma=1/2$ (Local) & 0.7700 [0.7688, 0.7712] & 1.0000 [1.0000, 1.0000] & 0.6292 [0.6262, 0.6321] & 0.0037 [0.0037, 0.0037] \\
$\gamma=1/3$ (Local) & 0.7571 [0.7569, 0.7574] & 1.0000 [1.0000, 1.0000] & 0.5989 [0.5984, 0.5994] & 0.0031 [0.0031, 0.0031] \\
$\gamma=2/3$ (Local) & 0.7527 [0.7196, 0.7859] & 1.0000 [1.0000, 1.0000] & 0.6187 [0.5141, 0.7233] & 0.0037 [0.0037, 0.0038] \\
$\gamma=4/5$ (Local) & 0.6679 [0.6647, 0.6711] & 1.0000 [1.0000, 1.0000] & 0.3943 [0.3919, 0.3967] & 0.0041 [0.0040, 0.0041] \\
\bottomrule
\end{tabular}
}
\end{table}

\begin{table}[ht]
\centering
\caption{Final-slot metrics under different slot-time settings.}
\label{tab:final-metrics-eip7782}
\resizebox{\textwidth}{!}{
\begin{tabular}{lcccc}
\toprule
\textbf{Configuration} & $\boldsymbol{\mathrm{Gini}_{\mathrm{g}}}$& $\boldsymbol{\mathrm{LC}_{\mathrm{g}}}$ & $\boldsymbol{\mathrm{HHI}_{\mathrm{g}}}$ & $\boldsymbol{\mathrm{CV}_{\mathrm{g}}}$ \\
\midrule
$\Delta=12s$ (External) & 0.2558 [0.2551, 0.2566] & 2.0000 [2.0000, 2.0000] & 0.1829 [0.1822, 0.1835] & 0.0018 [0.0018, 0.0018] \\
$\Delta=6s$ (External) & 0.2540 [0.2535, 0.2546] & 2.0000 [2.0000, 2.0000] & 0.1814 [0.1809, 0.1818] & 0.0019 [0.0019, 0.0020] \\
$\Delta=12s$ (Local) & 0.7527 [0.7196, 0.7859] & 1.0000 [1.0000, 1.0000] & 0.6187 [0.5141, 0.7233] & 0.0037 [0.0037, 0.0038] \\
$\Delta=6s$ (Local) & 0.7289 [0.6985, 0.7593] & 1.0000 [1.0000, 1.0000] & 0.5486 [0.4537, 0.6434] & 0.0040 [0.0040, 0.0041] \\
\bottomrule
\end{tabular}
}
\end{table}

\begin{table}[ht]
\centering
\caption{Final-slot metrics under different $\sigma$ values in the log-normal latency model.}
\label{tab:final-metrics-sigmas}
\resizebox{\textwidth}{!}{
\begin{tabular}{lcccc}
\toprule
\textbf{Configuration} & $\boldsymbol{\mathrm{Gini}_{\mathrm{g}}}$& $\boldsymbol{\mathrm{LC}_{\mathrm{g}}}$ & $\boldsymbol{\mathrm{HHI}_{\mathrm{g}}}$ & $\boldsymbol{\mathrm{CV}_{\mathrm{g}}}$ \\
\midrule
$\sigma=0.2$ (External) & 0.8571 [0.8571, 0.8571] & 1.0000 [1.0000, 1.0000] & 1.0000 [1.0000, 1.0000] & 0.0036 [0.0036, 0.0036] \\
$\sigma=0.3$ (External) & 0.8571 [0.8571, 0.8571] & 1.0000 [1.0000, 1.0000] & 1.0000 [1.0000, 1.0000] & 0.0030 [0.0029, 0.0032] \\
$\sigma=0.4$ (External) & 0.8571 [0.8571, 0.8571] & 1.0000 [1.0000, 1.0000] & 1.0000 [1.0000, 1.0000] & 0.0030 [0.0030, 0.0030] \\
$\sigma=0.5$ (External) & 0.8571 [0.8571, 0.8571] & 1.0000 [1.0000, 1.0000] & 1.0000 [1.0000, 1.0000] & 0.0030 [0.0030, 0.0030] \\
$\sigma=0.6$ (External) & 0.8571 [0.8571, 0.8571] & 1.0000 [1.0000, 1.0000] & 1.0000 [1.0000, 1.0000] & 0.0030 [0.0030, 0.0030] \\
$\sigma=0.7$ (External) & 0.8571 [0.8571, 0.8571] & 1.0000 [1.0000, 1.0000] & 1.0000 [1.0000, 1.0000] & 0.0030 [0.0030, 0.0030] \\
$\sigma=0.8$ (External) & 0.8571 [0.8571, 0.8571] & 1.0000 [1.0000, 1.0000] & 1.0000 [1.0000, 1.0000] & 0.0029 [0.0029, 0.0029] \\
$\sigma=0.2$ (Local) & 0.8571 [0.8571, 0.8571] & 1.0000 [1.0000, 1.0000] & 1.0000 [1.0000, 1.0000] & 0.0037 [0.0036, 0.0039] \\
$\sigma=0.3$ (Local) & 0.8571 [0.8571, 0.8571] & 1.0000 [1.0000, 1.0000] & 1.0000 [1.0000, 1.0000] & 0.0037 [0.0037, 0.0037] \\
$\sigma=0.4$ (Local) & 0.8571 [0.8571, 0.8571] & 1.0000 [1.0000, 1.0000] & 1.0000 [1.0000, 1.0000] & 0.0037 [0.0037, 0.0037] \\
$\sigma=0.5$ (Local) & 0.8571 [0.8571, 0.8571] & 1.0000 [1.0000, 1.0000] & 1.0000 [1.0000, 1.0000] & 0.0037 [0.0037, 0.0037] \\
$\sigma=0.6$ (Local) & 0.8571 [0.8571, 0.8571] & 1.0000 [1.0000, 1.0000] & 1.0000 [1.0000, 1.0000] & 0.0037 [0.0037, 0.0037] \\
$\sigma=0.7$ (Local) & 0.8571 [0.8571, 0.8571] & 1.0000 [1.0000, 1.0000] & 1.0000 [1.0000, 1.0000] & 0.0037 [0.0037, 0.0037] \\
$\sigma=0.8$ (Local) & 0.8571 [0.8571, 0.8571] & 1.0000 [1.0000, 1.0000] & 1.0000 [1.0000, 1.0000] & 0.0037 [0.0037, 0.0037] \\
\bottomrule
\end{tabular}
}
\end{table}

\begin{table}[ht]
\centering
\caption{Final-slot metrics under different migration costs.}
\label{tab:final-metrics-cost}
\resizebox{\textwidth}{!}{
\begin{tabular}{lcccc}
\toprule
\textbf{Configuration} & $\boldsymbol{\mathrm{Gini}_{\mathrm{g}}}$& $\boldsymbol{\mathrm{LC}_{\mathrm{g}}}$ & $\boldsymbol{\mathrm{HHI}_{\mathrm{g}}}$ & $\boldsymbol{\mathrm{CV}_{\mathrm{g}}}$ \\
\midrule
$c = 0.0$ (External) & 0.8571 [0.8571, 0.8571] & 1.0000 [1.0000, 1.0000] & 1.0000 [1.0000, 1.0000] & 0.0030 [0.0030, 0.0030] \\
$c = 0.001$ (External) & 0.8354 [0.8237, 0.8472] & 1.0000 [1.0000, 1.0000] & 0.8752 [0.8284, 0.9220] & 0.0035 [0.0035, 0.0036] \\
$c = 0.002$ (External) & 0.2558 [0.2551, 0.2566] & 2.0000 [2.0000, 2.0000] & 0.1829 [0.1822, 0.1835] & 0.0018 [0.0018, 0.0018] \\
$c = 0.003$ (External) & 0.0194 [0.0194, 0.0194] & 3.0000 [3.0000, 3.0000] & 0.1431 [0.1431, 0.1431] & 0.0014 [0.0014, 0.0014] \\
$c = 0.0$ (Local) & 0.8571 [0.8571, 0.8571] & 1.0000 [1.0000, 1.0000] & 1.0000 [1.0000, 1.0000] & 0.0037 [0.0037, 0.0037] \\
$c = 0.001$ (Local) & 0.8571 [0.8571, 0.8571] & 1.0000 [1.0000, 1.0000] & 1.0000 [1.0000, 1.0000] & 0.0037 [0.0037, 0.0037] \\
$c = 0.002$ (Local) & 0.7527 [0.7196, 0.7859] & 1.0000 [1.0000, 1.0000] & 0.6187 [0.5141, 0.7233] & 0.0037 [0.0037, 0.0038] \\
$c = 0.003$ (Local) & 0.2282 [0.2279, 0.2286] & 2.0000 [2.0000, 2.0000] & 0.1738 [0.1738, 0.1739] & 0.0026 [0.0026, 0.0026] \\
\bottomrule
\end{tabular}
}
\end{table}

\end{document}